\documentclass[conference]{IEEEtran}

\usepackage{graphicx}
\usepackage{mathtools}
\usepackage{amsfonts}
\usepackage{subfigure}
\usepackage[ruled,linesnumbered]{algorithm2e}

\IEEEoverridecommandlockouts

\newtheorem{definition}{Definition}
\title{EPASAD: Ellipsoid decision boundary based Process-Aware Stealthy Attack Detector}

\author{\IEEEauthorblockN{
        Vikas Maurya,
Rachit Agarwal,
Saurabh Kumar, Sandeep Kumar Shukla
    }
    \IEEEauthorblockA{
        CSE, IIT Kanpur, India\\
        Email: \{vikasmr, rachitag, skmtr, sandeeps\}@cse.iitk.ac.in
    }
}

\begin{document}

\sloppy
\raggedbottom
\pagenumbering{gobble}
\maketitle

\begin{abstract}
Due to the importance of Critical Infrastructure (CI) in a nation's economy, they have been lucrative targets for cyber attackers. These critical infrastructures are usually Cyber-Physical Systems (CPS) such as power grids, water, and sewage treatment facilities, oil and gas pipelines, etc. In recent times, these systems have suffered from cyber attacks numerous times. Researchers have been developing cyber security solutions for CIs to avoid lasting damages. According to standard frameworks, cyber security based on identification, protection, detection, response, and recovery are at the core of these research. Detection of an ongoing attack that escapes standard protection such as firewall, anti-virus, and host/network intrusion detection has gained importance as such attacks eventually affect the physical dynamics of the system. Therefore, anomaly detection in physical dynamics proves an effective means to implement defense-in-depth. PASAD is one example of anomaly detection in the sensor/actuator data, representing such systems' physical dynamics. We present EPASAD, which improves the detection technique used in PASAD to detect these micro-stealthy attacks, as our experiments show that PASAD's spherical boundary-based detection fails to detect. Our method EPASAD overcomes this by using Ellipsoid boundaries, thereby tightening the boundaries in various dimensions, whereas a spherical boundary treats all dimensions equally. We validate EPASAD using the dataset produced by the TE-process simulator and the C-town datasets. The results show that EPASAD improves PASAD's average recall by 5.8\% and 9.5\% for the two datasets, respectively.

\end{abstract}


\section{Introduction}
Critical infrastructures (CIs) are mostly Cyber-Physical Systems (CPS) with few exceptions (such as telecommunication, financial services, and Agriculture) that facilitate and boost societal and economical operations. Some examples of CIs include infrastructure supporting supply of natural gas, water treatment and supply, electricity generation and renewable energy, food production and distribution, transportation, healthcare, and goods and services. The architecture of a CI is layered- an industrial control system (ICS - also known as cyber-physical systems (CPS)), Supervisory Control and Data Acquisition systems (SCADA), and Process Control Systems (PCS or Distributed Control Systems (DCS)) monitor and control the infrastructure~\cite{cardenas2011attacks}. These high-level designs of supervisory systems are often networked with Programmable Logic Controllers (PLCs). PLCs are industrial computational devices coupled with sensors and actuators to control physical processes by communicating usually with SCADA. The SCADA system is comprised of numerous intrusion detection systems (IDSs) that monitor physical processes or network data generated by sensors and actuators on a regular basis and generate an alarm if the system behaves abnormally.


Minor damage to a CI may lead to catastrophe and significantly impacts public safety, economy, and daily life demands. With the rise of the Internet and connected things, CIs have become more vulnerable to cyber-attacks. A state's vested interests further escalate this. In the past, there have been numerous cases where cyber-criminals successfully infiltrated CIs. For example, an attack on the Iranian power plant in 2009 was conducted using Stuxnet~\cite{falliere2011w32} malware. Other examples of such attacks include (\textit{i}) attack on a German steel mill in 2014~\cite{lee2014german} that was conducted using spear-phishing through mails, (\textit{ii}) attack on the Ukrainian power grid in 2015 which was conducted using spear-phishing via Microsoft doc file affecting $\approx$225,000 customers~\cite{case2016analysis}, and (\textit{iii}) attack on a Saudi petroleum refinery in 2017 using TRITON malware caused the refinery to shut down its operations~\cite{di2018triton}. Besides these, there are numerous recent attempts reported by the center for strategic and international studies (CSIS). These include attacks on Indian nuclear plant in 2019, Israel water treatment plant in 2020, and oil and natural gas pipeline companies in USA in 2021~\cite{CSIS}.

Thus, the question that motivates us is \textit{how we can secure CIs from such attacks?} Multiple methods are used to answer this and secure CIs. These methods include securing network architecture by adhering to the policies such as network segmentation and segregation, the use of boundary protection devices, and firewall filters between each network segment~\cite{stouffer2015guide}. However, network security is constantly being breached due to exploitation of vulnerabilities that also include zero-day attacks. Assuming that network security is foolproof and no attacker will break it to cause harm to the ICS is not correct. Only by bypassing the network security attackers do not harm the CI until they perform any malicious activity. When an attacker performs any malicious activity on the CI, it gets reflected in the physical processes~\cite{zheng2015perceptions}. The sensors and control behaviour associated with attack-targeted devices start to show structural changes in their normal behaviour. Usually it happens in \textbf{direct damage attack} (DDA). Such structural changes can be identified to detect an attack.
However, an attacker can hide their manipulation within the noise margin. These attacks are known as \textbf{stealthy attacks} (SA). Such attacks are likely to produce a cascading effect due to the interaction of control loops, eventually causing the control system to fail. Further, an attacker can reduce the impact of SA in such way the sensor produced abnormal structural changes do not deviate much from the normal behaviour. We call such attacks as \textbf{micro-stealthy attacks} (MSA) (cf. Section \ref{fig:MSA}). These attacks are extremely difficult to detect and evade current state-of-the-art detection techniques. The MSA does not interrupt or fail the control system but slowly degrades the system causing huge losses in terms of money and raw material over an extended period. 
In this paper, we develop a novel IDS framework whose objective is to detect the most challenging attack category MSA, and quickly detect the SA and DDA to save the CIs from lasting damage. 

A process-level intrusion detection system (IDS) continuously monitors the physical process of ICSs. It is deployed over SCADA, whose goal is to detect any abnormal structural changes in the physical process behaviour. State-of-the-art approaches categorize process level IDS in two categories: univariate (independent IDS for each sensor) and multivariate (Single IDS for multiple sensor). Among the many popular IDS-based solutions (discussed in Section~\ref{sec:related}), \textbf{PASAD~\cite{aoudi2018truth} (cf. Section~\ref{subsec:sumPASAD}) is the most promising framework}. Yet, \textbf{PASAD suffers from same drawbacks as mentioned earlier} (cf. Section~\ref{subsec:attackmodel}). It fails to detect the \textbf{micro-stealthy attacks} (MSAs), and is delayed in detecting stealthy attacks (SAs) and direct damage attacks (DDAs). We consider PASAD as a baseline for the validation of our approach.

An efficient and realistic process level IDS must fulfill following objectives: \textit{(i)} be capable of detecting an attack before lasting damage, \textit{(ii)} be secure against evasion attack, \textit{(iii)} work under noisy environment, \textit{(iv)} be realistic to build and deploy, \textit{(v)} have less computational overhead and produce the result for streaming data quickly, and \textit{(vi)} have lower false alarm rate. Motivated by this, we present an efficient and realistic process level IDS solution called \textbf{EPASAD} (Ellipsoid decision boundary based Process-Aware Stealthy Attack Detector), which addresses above-mentioned objectives. EPASAD is a univariate process-level IDS based on Singular Spectrum Analysis (SSA)~\cite{broomhead1986extracting, elsner2013singular, golyandina2014basic, golyandina2001analysis,golyandina2013singular, hassani2010brief, vautard1989singular}, a time series analysis tool. EPASAD is designed to detect any structural changes in a sensor behaviour caused due to the presence of an adversary. EPASAD projects raw sensor data into a noise-free lower-dimensional signal subspace to cluster normal data. It uses the distribution of the clustered data to learn an efficient and uniformly tight decision boundary. EPASAD envelops the signal subspace within an ellipsoid decision boundary. After learning the decision boundary, any sensor datum that falls outside is considered abnormal, and an alarm is raised. From the attacker's perspective, in PASAD, it is easier to compromise huge redundant normal space within the decision-boundary and easier to determine the radius using any one projected dimension. However, EPASAD provides a tighter boundary with respective radii in each dimension, necessitating more effort for attackers to determine ellipsoid parameters in each projected dimension to stay under the radar. This is the novelty that we bring over the existing state-of-the-art approaches.

We consider various attack scenarios present in two datasets: TE-process and C-town dataset to validate EPASAD. We use TE chemical process simulator to generate two MSAs, three SAs, and two DDA scenarios with the motivation to simulate realistic situations. Our results show that EPASAD successfully detects the MSAs and quickly detects the stealthy and the DDAs when compared to the baseline method PASAD. Further, we consider C-town network datasets to validate our approach on a much larger dataset. Using the dataset, we validate EPASAD using 14 different attacks scenarios that happen over a testing period of 9-months. Each attack present in the C-town dataset is sandwiched between a long-duration normal operation. Testing for such an extended period validates our framework for a realistic scenario. We show that EPASAD is capable of detecting all 14 attacks present in the C-town dataset with a low false alarms rate of 3.7\%. Compared to PASAD, EPASAD shows a significant improvement over each attack scenarios. Over PASAD, EPASAD improves the overall recall for all the sensors in the system operating under MSAs, SAs, and DDAs present in the TE-dataset from 7.5\% to 17.3\%, 50.3\% to 54.2\%, and 46.2\% to 51.0\%, respectively. Similarly, in the C-town dataset, EPASAD improves the overall recall from 54.8\% to 64.3\% for all the 14 attacks present in it. When an attacker attacks a CPS, the behavioural change to anomalous state takes time. But the training data is labelled as ``attack'' as soon as the attacker engages. This is why the low accuracy appears in both the PASAD and EPASAD.  In such low accuracy scenarios, an improvement of even 3.9\% (the minimum average gain among all scenarios mentioned above) might appear small from an absolute perspective. From a relative perspective, it is a significant improvement.

In summary, the major contributions of our work are:
\begin{itemize}
\item We introduce an attack scenario called \textbf{\textit{Micro-Stealthy Attack (MSA)}}, which although existed but was not studied before and posed detection challenges for current state-of-the-art approaches (cf. Section~\ref{subsec:attackmodel}). 

\item Our framework called \textbf{\textit{EPASAD}} provides an efficient and realistic process-level univariate IDS for securing CIs. EPASAD continuously monitors the data stream consisting of sensor measurements for detecting tiny structural changes in the normal behaviour hidden within the noise margin.

\item We validate EPASAD on MSA and find that EPASAD efficiently detects them. Further, EPASAD \textit{\textbf{significantly improves PASAD}} without any additional computation overhead. We compare EPASAD with PASAD, using multiple attack scenarios present in the TE-process and C-town dataset (cf. Section~\ref{sec:experiment}).
\end{itemize}

The rest of the paper is organised as follows: first we discuss the required concepts that form a background knowledge needed to understand EPASAD in Section~\ref{sec:background}. Then, in Section~\ref{sec:meth}, we describe the attack model that forms the motivation behind the proposal of the EPASAD framework. In the Section~\ref{subsec:EPASAD}, we present our proposed framework EPASAD and provide detail of its training, online testing process, and computation cost analysis. In Section~\ref{sec:data}, we describe the generated and the existing datasets used for validation. In Section~\ref{sec:experiment}, we experimentally validate our method under three subsections and report our results. In Section~\ref{sec:related}, we discuss the related works, mainly highlighting the process-level IDS. Finally, in Section~\ref{sec:discandconclusion} we conclude our paper along with an in-depth discussion.

\begin{table}
\caption{Notations and their description}\label{tab:parameters}
\centering
\begin{tabular}{|l|l|}
\hline
 Notation  & Description\\
\hline \hline
$\mathbb{R}$ & Set of Real numbers\\
$\mathbb{I}$ & Set of Integers\\
$m_i$ & $i^{th}$ Measurement\\
$\mathbf{M}$ & Trajectory Matrix of size $L\times K$\\
$m$ & $L-$length lagged vector \\
$M_i$ & A specific lagged vector of length $L$, $i^{th}$ column vector\\
& of $\mathbf{M}$ or test subsequence for $i^{th}$ measurement\\
$c$ &  Centroid vector in $\mathbb{R}^L$ \\
$\mathbf{P}$ & Projection matrix\\
$\mathbf{U}$ & Eigen matrix\\
$U_i$ & $i^{th}$ Eigenvector\\
$\mathbf{X}$ & A signal subspace matrix of size $R\times K'$\\
$X_i$ & A specific $R-$length lagged vector in $\mathbb{R}^R$, $i^{th}$ column vector\\
& of $\mathbf{X}$ or projected test subsequence for $i^{th}$ measurement\\
$x$ &  A $R-$length lagged vector in signal subspace\\
$w$ &  A weight vector in $\mathbb{R}^R$\\
$\hat{c}$ & Centroid vector in $\mathbb{R}^R$ \\
$\mathcal{D}_t$ & Departure score at timestamp $t$\\
$\theta_p$ & Threshold of PASAD\\
$\theta_e$ & Threshold of EPASAD\\
$\delta_f(x)$ & Tightness of decision boundary $f(x)$ at a point $x$\\
$N$ & Length of training subsequence\\
$N'$ & Length of training+validation subsequence\\
$L$ & Lag parameter in $\mathbb{I}$\\
$R$ & Dimensionality of signal subspace parameter\\
$\epsilon$ & Slag-value parameter\\
$\prod(w)$ & Product of elements of vector $w$\\
\hline
\end{tabular}
\end{table}

\section{Background}\label{sec:background}
In this section, we discuss the techniques and the concepts that are useful for this work. 

\subsection{Singular Spectral Analysis (SSA)} \label{subsec:SSA}

SSA is a non-parametric model-free time series analysis tool with a wide range of applications~\cite{broomhead1986extracting, elsner2013singular, vautard1989singular,golyandina2001analysis,mohammad2011comparing}, including IDS~\cite{aoudi2018truth, terai2018intrusion, dong2017anomaly, moskvina2003change,golyandina2001analysis,mohammad2011comparing}. SSA can robustly recover the deterministic pattern of a time series even in the presence of noise. Such aspects of SSA enable us to use it to analyse the noise-free structure of a time series. SSA is also used to identify structural changes in a time series data by learning a projection matrix $\mathbf{P}$ that projects a real-valued noisy subseries into a noise-free signal subspace. However, to do so, only two steps of SSA are sufficient. As our focus is to identify structural changes in normal sensor measurements, here, we only explain these two steps. Note that a summary of the notations/symbols used in this paper are listed in the Table~\ref{tab:parameters}. The two steps are:

\textbf{Step 1: (Embedding)} This step maps a univariate time series into a trajectory matrix. Let $\mathcal{T}$=$\{m_1,m_2,\cdots,m_N\}$ be a univariate time series of length $N$ where $m_i\in\mathbb{R}$ is a sensor's measurement collected at the $i^{th}$ timestamp. Let $L\in\mathbb{I}$ where $1<L<N/2$ be called as lag or window-length and $K$=$N-L+1$. The SSA arranges the time series $\mathcal{T}$ in the form of a trajectory matrix $\mathbf{M}$ of dimension $L\times K$.
\[
\mathbf{M}=
\begin{bmatrix}
m_1 & m_2 & ... & m_K \\ 
m_2 & m_3 & ... & m_{K+1} \\ 
. & . &   & . \\ 
. & . &   & . \\ 
m_L & m_{L+1} & ... & m_N
\end{bmatrix}
\label{matrix:hanckelmatrix}
\]

A column vector of $\mathbf{M}$ is called the lagged vector where the $i^{th}$ $(1\leq i \leq K)$ lagged vector ($M_i$) is defined by $M_i$=$[m_i,m_{i+1},\cdots,m_{i+L-1}]^T$.

\textbf{Step 2: Singular Value decomposition (SVD)} In this step, SVD of $\mathbf{M}$ is done by using the following four steps: \textit{(i)} compute a lagged co-variance matrix $\mathbf{MM}^T$ of dimension $L\times L$, \textit{(ii)} compute the eigenvalues denoted by $\lambda_1$,$\lambda_2$, $\cdots$,$\lambda_L$ and the corresponding eigenvectors denoted by $U_1$, $U_2$,$\cdots$, $U_L$, which are arranged according to decreasing magnitude of eigenvalues, \textit{(iii)} orthonormalise the eigenvectors, and \textit{(iv)} pick $R$ leading eigenvectors to form eigen matrix $\mathbf{U}$ of dimension $L\times R$, i.e., $\mathbf{U}$=$[U_1, U_2,\cdots, U_R]$. Ignoring the minor and keeping the leading eigenvectors in the eigen matrix $\mathbf{U}$ eliminates the noise and retains the deterministic behaviour of a signal subseries. The set of eigenvectors $\{U_1,U_2,\cdots,U_R\}$ are linearly independent, spanning an $R-$dimensional subspace in $\mathbb{R}^L$ (length of vectors in the $R-$dimensional subspace is $L$) called \textbf{signal subspace}. There exists $\mathbf{P}$=$\mathbf{U}(\mathbf{U}^T\mathbf{U})^{-1}\mathbf{U}^T$=$\mathbf{U}\mathbf{U}^T$ (since $\mathbf{U}$ is an orthonormal matrix, then $\mathbf{U}^T\mathbf{U}$=$\mathbf{I}$) that projects a lagged vector from $L-$dimensional real space to the signal subspace. Let $m\in\mathbb{R}^L$ be a lagged vector, then the projection of $m$, i.e., $\mathbf{P}m\in \mathbb{R}^L$, be a noise-free vector in signal subspace. Note that the notation $m$ is an $L-$length variable lagged vector while $M_i$ is a constant representing $i^{th}$ column vector of matrix $\mathbf{M}$.
\subsection{PASAD:} \label{subsec:sumPASAD}
In~\cite{aoudi2018truth}, the authors describe PASAD, a process-level, univariate, and anomaly-based IDS that monitors ICS process activity in real-time to determine whether the system is operating normally or abnormally. The motivation behind is to detect any aberrant structural change in the physical process to detect stealthy and direct damage attacks.

PASAD leverages from SSA to learn $\mathbf{P}$=$\mathbf{U}\mathbf{U}^T$. To reduce the computational overhead of PASAD, in~\cite{aoudi2018truth}, the authors proved that an $L-$dimensional lagged vector $m$ projected by $\mathbf{P}$=$\mathbf{U}^T$ in $\mathbb{R}^R$ preserves the Euclidean distance projected by $\mathbf{P}$=$\mathbf{U}\mathbf{U}^T$ in $\mathbb{R}^L$, i.e.,   $||\mathbf{U}\mathbf{U}^Tv||$=$||\mathbf{U}^Tv||$. The $\mathbf{P}$=$\mathbf{U}^T$ captures the deterministic behaviour of the physical process by projecting an $L-$dimensional normal subseries onto a lower $R-$dimensional signal subspace. PASAD computes the squared Euclidean from centroid in $R-$dimensional signal space for each streaming test lagged vectors $M_i$ ($i>N$) called departure score ($\mathcal{D}_i$) to detect the attack-induced structural changes in the normal behaviour. $\mathcal{D}_i$ is defined via Equation~\ref{eq:1departurescorePASAD}, where $\hat{c}$=$\mathbf{U}^Tc$ and $c$ is the mean of column vectors of $\mathbf{X}$, i.e., $c$=$\sum_{i=1}^{K}X_i$.
\begin{equation}\label{eq:1departurescorePASAD}
        \mathcal{D}_i=||\hat{c}-\mathbf{U}^TM_i||^2
\end{equation}
The projection of the normal subseries forms a dense cluster which is closer to the center. While an abnormal subseries is forced to be projected far away from the center of a normal cluster ($\hat{c}$) thereby having a higher departure score. If the departure score crosses certain threshold $\theta_p$, i.e., if $||\hat{c}-\mathbf{U}^Tv_i||^2>\theta_p$, an attack alarm is triggered.

To compute $\theta_p$, PASAD computes departure scores on training measurements and few extended measurements collected during normal operation. The extended measurements are called validation dataset. PASAD sets $\theta_p=\max_{\forall i}(\mathcal{D}_i)$. As a result, PASAD forms an $\mathbf{n-}$\textbf{spherical decision boundary} (an $n-$sphere is a generalised form of a sphere in the n-dimensions) in $R-$dimensional signal subspace. The radius of the $n-$sphere is $\sqrt{\theta_p}$ which is the distance of the farthest normal point from center ($\hat{c}$) in the signal subspace.

PASAD is a lightweight IDS suitable for deploying on limited hardware resources. PASAD's most computationally intensive step is to project the $L-$dimensional vector into an $R-$dimensional signal space which is an $R\times L$ dimensional matrix to $L$ dimensional vector multiplication. As a result, the computational complexity of the PASAD is $O(RL)$.

\section{Attack Model} \label{sec:meth}
In this section, we discuss the attack model that encompasses the motivation for developing EPASAD along with necessary definitions. 

\begin{definition}[Normal cluster] \label{defn:normalcluster}
Set of normal points (column vectors of $\mathbf{X}$ in Equation~\ref{eq:2signalnormalcluster}) in signal subspace collected by projecting the measurements when there was no attack (also referred as normal measurements).
\end{definition}

\begin{definition}[Decision boundary] \label{defn:decisionboundary}
A non-linear function $f(x)$ encloses the normal cluster and separates the projection of the measurements captured under attack (also referred to as attack measurements) and normal operations.
\end{definition}

\begin{definition}[Tightness of decision boundary] \label{defn:tightness}
Let $x_1$ and $x_2$ be two points on a decision boundary $f(x)$, points $x'_1$ and $x'_2$ be the nearest (shortest Euclidean distance) points of the normal cluster from $x_1$ and $x_2$, respectively. The distance between $x_1$ and $x'_1$ be $\delta_f(x_1)$=$||x_1-x'_1||$ defined as tightness of the decision boundary $f(x)$ at $x_1$, similarly for $x_2$. If $\delta_f(x_1)<\delta_f(x_2)$, then the decision boundary $f(x)$ is tighter at $x_1$ in comparison to $x_2$. In other words, $f(x)$ is loose at $x_2$ than $x_1$. 
\end{definition}

\begin{definition}[Uniformly tight decision boundary] \label{defn:uniform}
Let $f(x)$ and $g(y)$ be the two decision boundaries, and if $|max(\delta_f(x))-min(\delta_f(x))|<|max(\delta_g(y))-min(\delta_g(y))|$, then we call $f(x)$ is more uniformly tight decision boundary than $g(y)$.
\end{definition}

\subsection{Direct damage attack (DDA):} \label{subsec:DDAbackground}

A DDA is a conventional attacking approach where an attacker does not hide their malicious activities in the physical process. A DDA attacker's goal is to damage the devices and eventually interrupt the process. Here, the attacker tries to accomplish his harmful goals before being detected and make CI operate abnormally. These attacks are trivial to be detected, but any delay in their detection causes severe consequences for a CI. An efficient IDS aims to detect abnormal behaviour induced by such attacks at the initial stages to save CIs from lasting damage.

\subsection{Stealthy attack (SA):} \label{subsec:SAbackground}

In~\cite{feng2017multi}, the authors argued that in a noisy environment, a strategic attacker benefits from inflicting a substantial perturbation on the system state. The attack escapes the detection by failure and anomaly detectors as they do not consider noise. Strategic attackers' goal is to cause slow damaging perturbations in the physical process while being undetected for an extended period. Such attacks are likely to produce a cascading effect due to the interaction of control loops, eventually causing the control system to fail. Sometimes a strategic attacker may mask their attack so that the reflected anomaly in physical process variables remains within the noise level; the noise can be manufactured intentionally by the attacker or naturally by the system. Attacks that hide their manipulation within noise margin are known as SAs.



\begin{figure}
    \vspace{-0.75cm}
    \centering
	\subfigure{
	    \includegraphics[width=0.95\columnwidth]{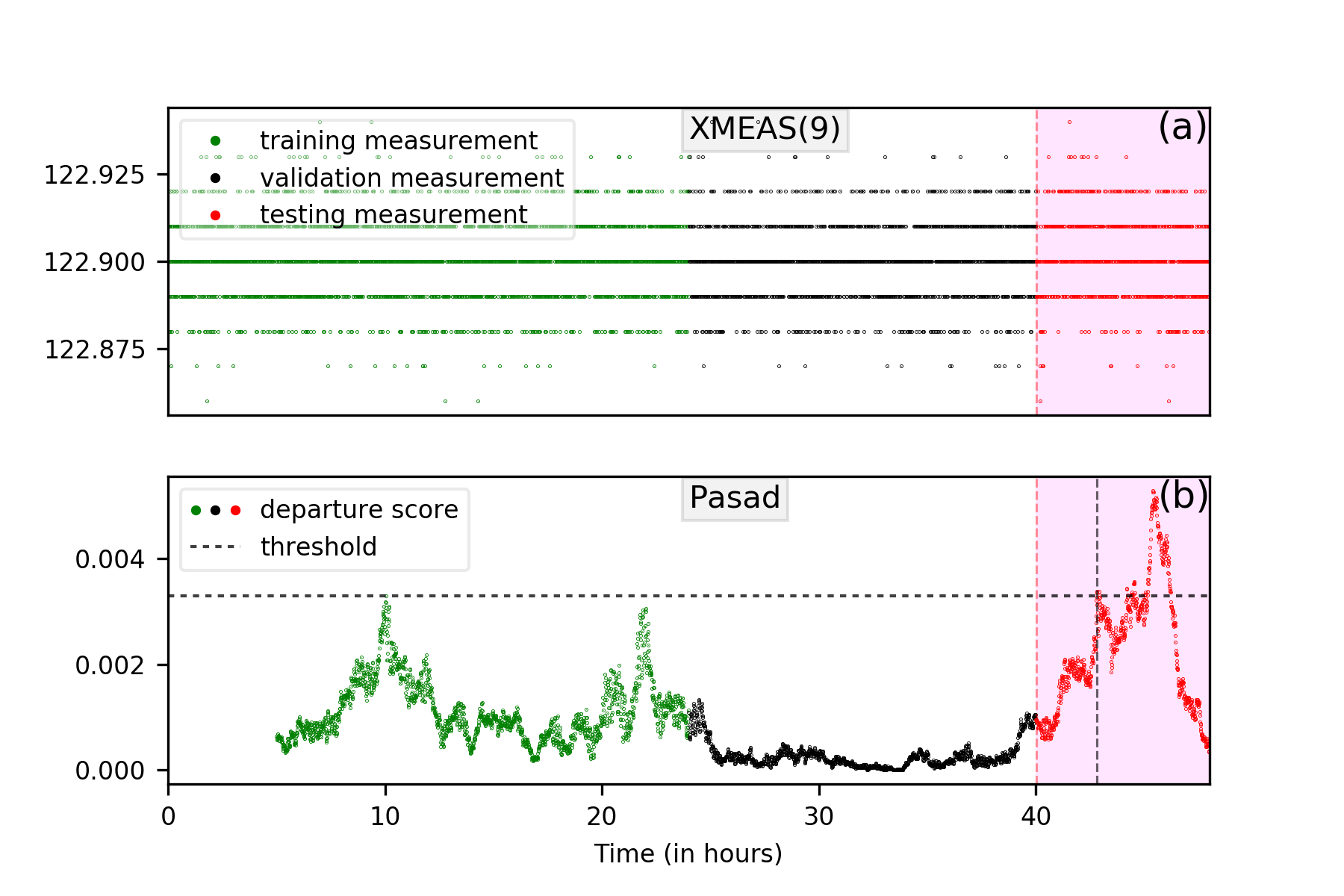}
	    
	}\vspace{-0.5cm}\\
	\subfigure{\vspace{-1cm}
	    \includegraphics[width=0.95\columnwidth]{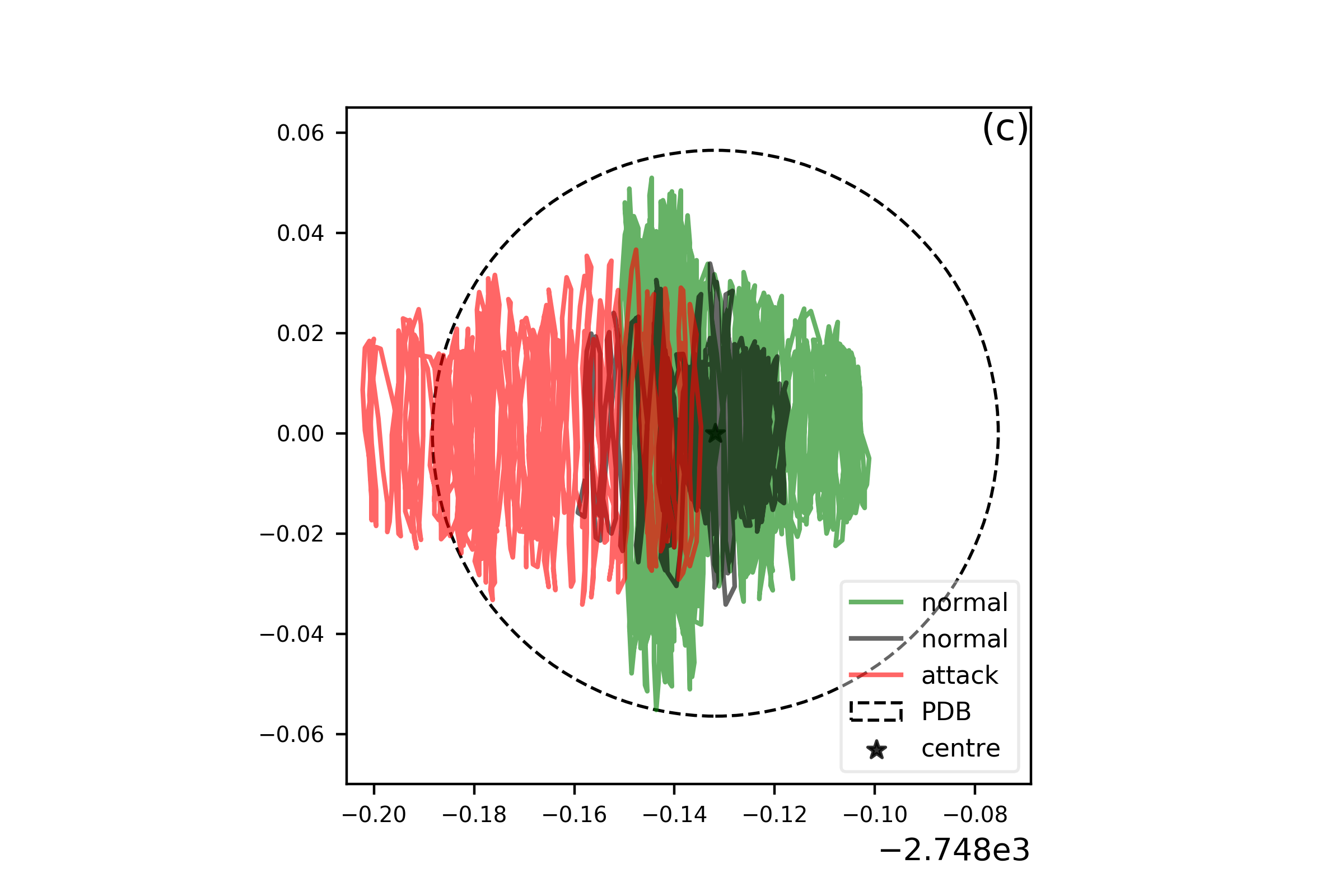}
	   }\vspace{-.5cm}
	\caption{We demonstrate a stealthy attack scenario on a reactor's temperature sensor (XMEAS(9)). Here, PASAD framework is delayed in detecting the attack because of the projection of attacked measurements towards the loose side of decision boundary. Subfigure (a) shows the sensor-generated measurements. The green and black measurements are the normal measurements used for training and validation, respectively, while the red measurements are captured under a stealthy attack (SA3). Subfigure (b) represents the departure score of corresponding measurements generated by PASAD frameworks. Subfigure (c) demonstrates the projections of each normal and attack measurement on the signal subspace (we consider a $2-$dimensional signal subspace for better visualization) and the PASAD's decision boundaries (PDB).}
	\label{fig:attackmodel1}\vspace{-0.4cm}
\end{figure}

\begin{figure}
    \vspace{-0.75cm}
	\subfigure{\vspace{-1cm}
	    \includegraphics[width=0.95\columnwidth]{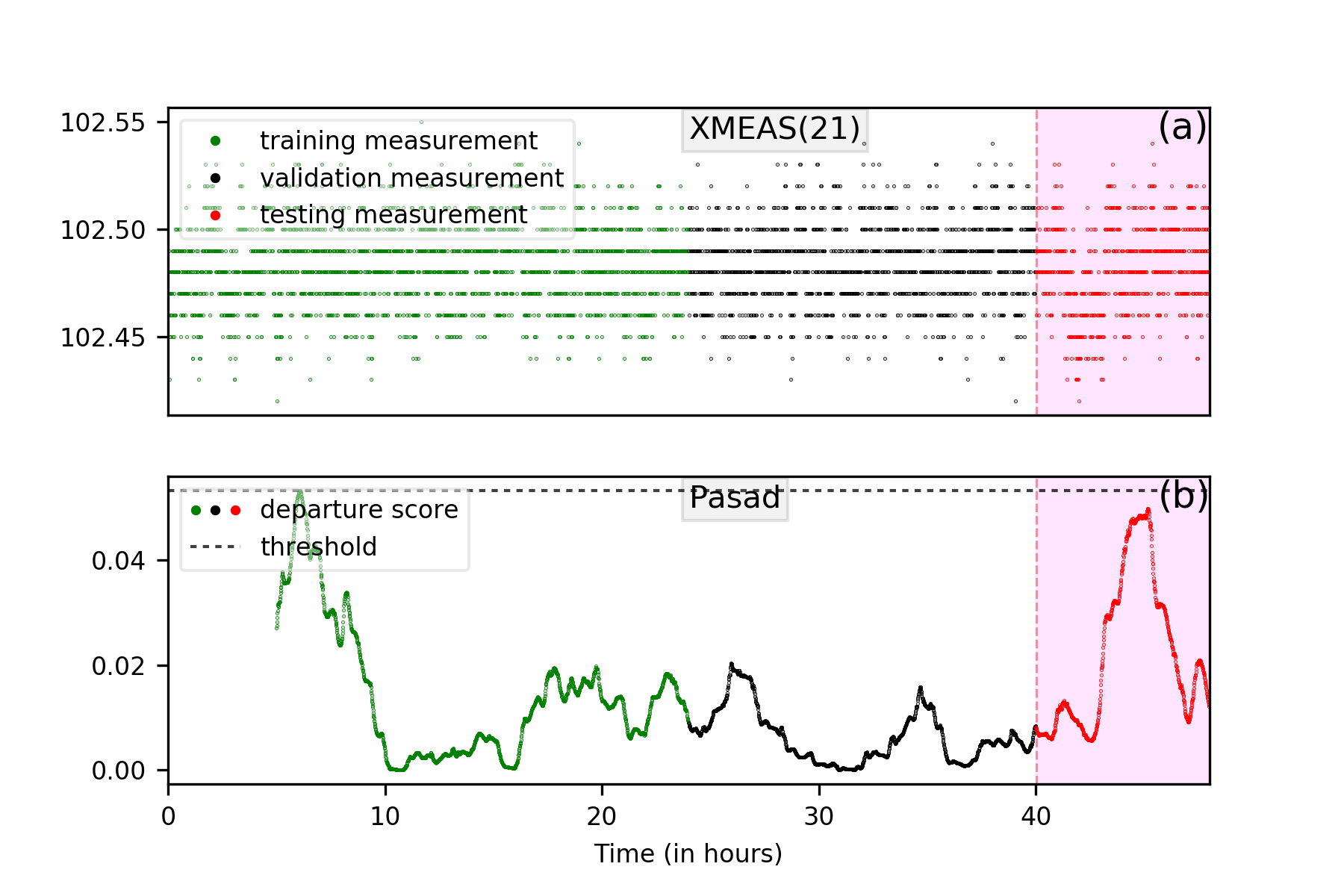}
	}\vspace{-0.5cm}\\
	\subfigure{
	\includegraphics[width=0.95\columnwidth]{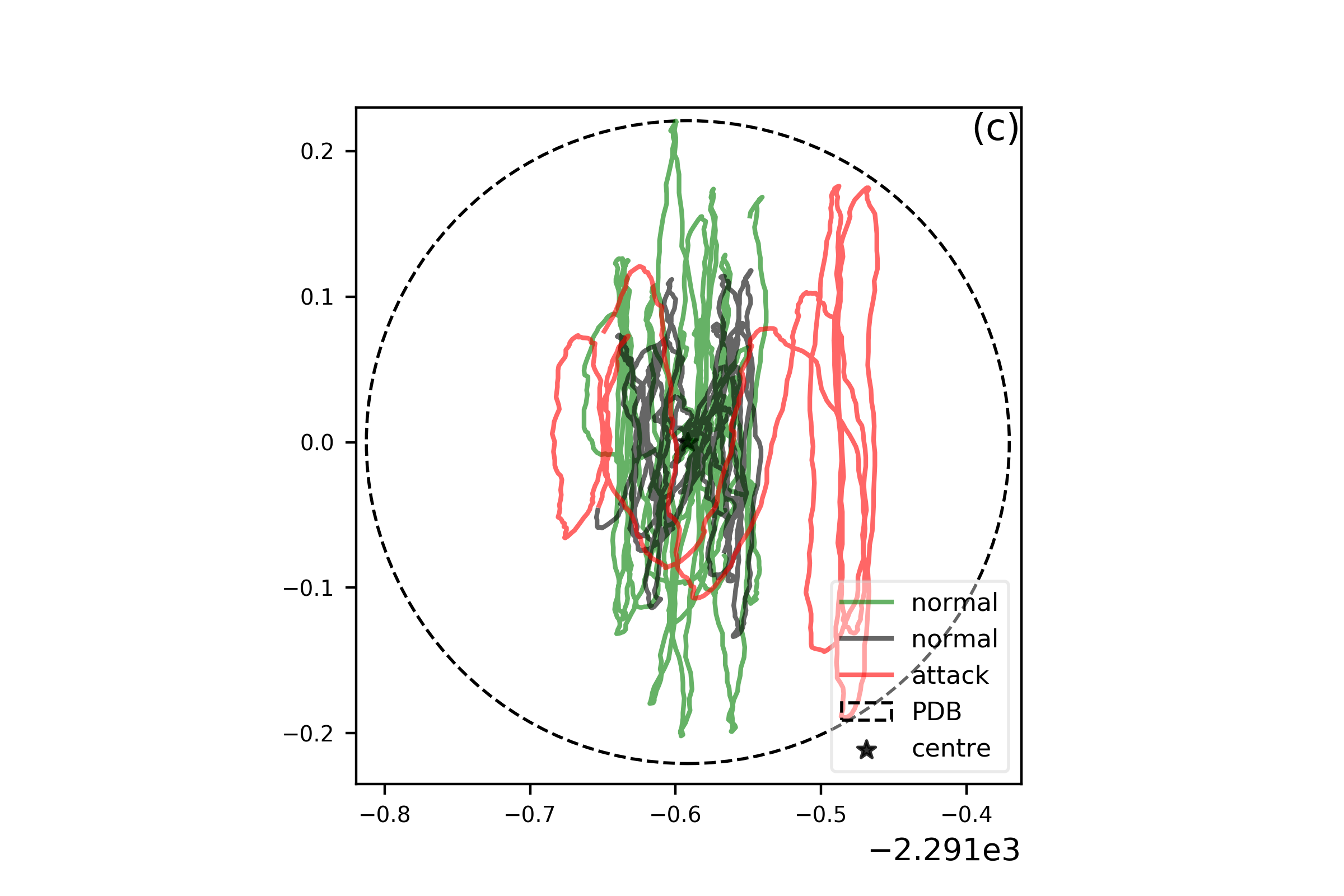}
	}\vspace{-.5cm}
	\caption{We demonstrate an MSA scenario where PASAD framework fails to detect the attack because of the attack's projection towards the loose side. Subfigure (a) shows measurements generated by the reactor's cooling water outlet temperature sensor (XMEAS(21). The red measurements are captured under a micro-stealthy attack (MSA1). Note that all other aspects and subfigures have same definition as Figure~\ref{fig:attackmodel1}.
	}
	\label{fig:attackmodel2}\vspace{-0.4cm}
\end{figure}

\subsection{Micro Stealthy attack (MSA)} \label{subsec:attackmodel} 

There have been several attack incidents where attackers compromised CIs by either installing malware, misusing the resources, making user compromise, performing Denial-of-Service (DoS) attacks, making root compromise, and performing social engineering attacks~\cite{kovacevic2015cyber}. An attacker's abnormal activities cause structural changes in the physical process. As attackers aim to cause maximum damage without being detected, a smart attacker hides the abnormalities by controlling the manipulations. In this paper, we do not focus on conducting an attack but on detecting any structural change in the normal behaviour caused by attacker activities. In~\cite{aoudi2018truth}, the authors present PASAD that detects such structural changes. However, PASAD has drawbacks. An attacker can evade PASAD by controlling the structural changes. Since PASAD envelops the $R-$dimensional signal subspace in an n-spherical decision boundary, one side is tight enough while the remaining are loose. There is a high probability that an attack-induced abnormal subseries get projected toward the loose side, or an attacker targets the abnormal projection towards the loosest side to hide the maximum abnormal manipulations. The projection towards the loose side causes serious issues such as delay in detecting the SAs, DDAs, and inability to detected some low-intensity attacks. We refer to such low-intensity SAs as \textbf{micro-stealthy attacks (MSAs)}.

In Figure~\ref{fig:attackmodel1}, we demonstrate the problem caused by a non-uniform loose decision boundary. Figure~\ref{fig:attackmodel1}(a) shows a time series of the reactor's temperature captured by the sensor XMEAS(9) of TE-process, initially under normal (green and black measurements) operation and ended with a SAs (red measurements) operation. We use the measurements under normal operation (green measurements) to determine $\mathbf{P}$. The other points under normal conditions (black measurements) determines the decision boundary. Finally, we test the model using the measurement (red measurements) captured under attack. Figure~\ref{fig:attackmodel1}(b) demonstrates the departure score of each sensor measurements computed by PASAD framework. 

We further demonstrate the projections of each normal and attack measurement on a $2-$dimensional signal subspace (cf. Figure~\ref{fig:attackmodel1}(c)) for better visualisation. In this $2$-dimensional signal subspace, the red points (attack subsequence projections) are projected far enough away from the green point's cluster. Since the abnormal projections are towards the loose side, it takes a long time to cross the spherical decision boundary of PASAD, causing a delay in detecting the SA. Thus, a question arises: \textit{What if a strategic attacker slightly reduces the SA's impact and attempts an MSA, never to cross the decision boundary?} PASAD will not detect the MSA attack that silently damages the CI and wastes valuable resources. We demonstrate such MSA attack scenario using Figure~\ref{fig:attackmodel2}. Figure~\ref{fig:attackmodel2}(a) represents measurements generated by sensor XMEAS(21) (represents reactor's cooling water outlet temperature) captured under an MSA scenario (cf. Section~\ref{subsec:TEdata} - MSA1). Here, the attacker manipulates the purge valve (XMV6) slightly higher than normal with the objective of wasting the reactor's gases. Figure~\ref{fig:attackmodel2}(c) shows that the attack-induced manipulated measurements are projected far enough from the normal cluster. Since the projections are toward the loose side and the impact of the attack is not that high to cross the decision boundary, the departure score of PASAD has never crossed the threshold (cf. Figure~\ref{fig:attackmodel2}(b)) and fails to detect the attacks reflected in XMEAS(21). Thus, we introduce EPASAD with a motivation to quickly detect the MSA, SA, and DDA.

\section{Proposed Framework: EPASAD}\label{subsec:EPASAD} 

EPASAD is a process-level, univariate, and anomaly-based IDS framework that monitors ICS process activity in real-time to determine whether the system is under normal or abnormal operation. Due to SSA's noise cancellation property, EPASAD works even in a noisy environment.

EPASAD collects the set of normal subseries on the signal subspace and envelops it within an efficient decision boundary. The subseries captured under normal operation follow certain oscillation and trend structures, projecting a set of normal subseries that forms a dense cluster of normal points. While an abnormal subsequence that has some structural manipulations get projected far from the normal cluster. An attack alarm is triggered if the projection surpasses the decision boundary.

\begin{figure}
\centering
	\vspace{-0.75cm}
	\subfigure{
	    \includegraphics[width=0.95\columnwidth]{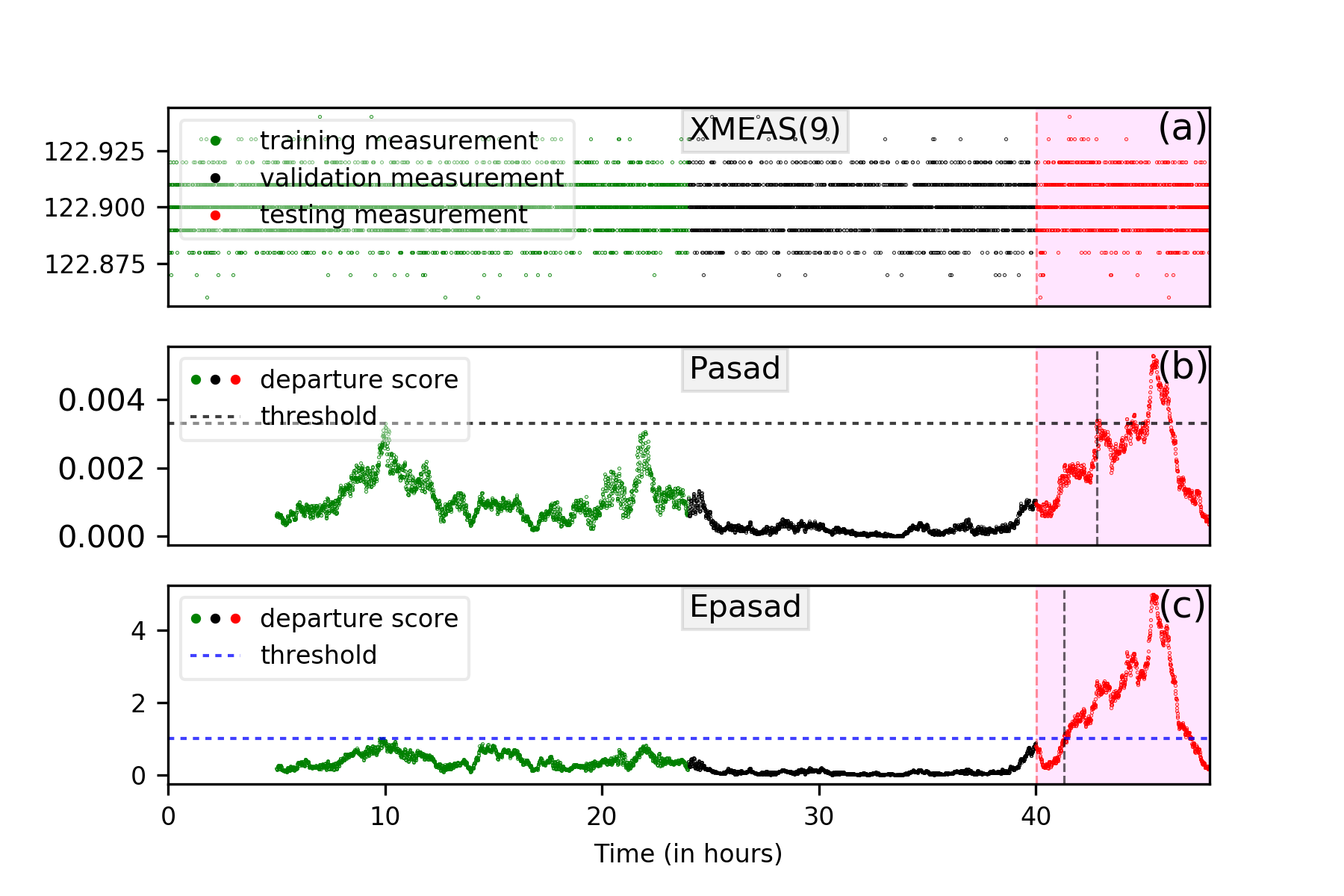}
    }\vspace{-0.5cm}\\
	\subfigure{
	    \includegraphics[width=0.95\columnwidth]{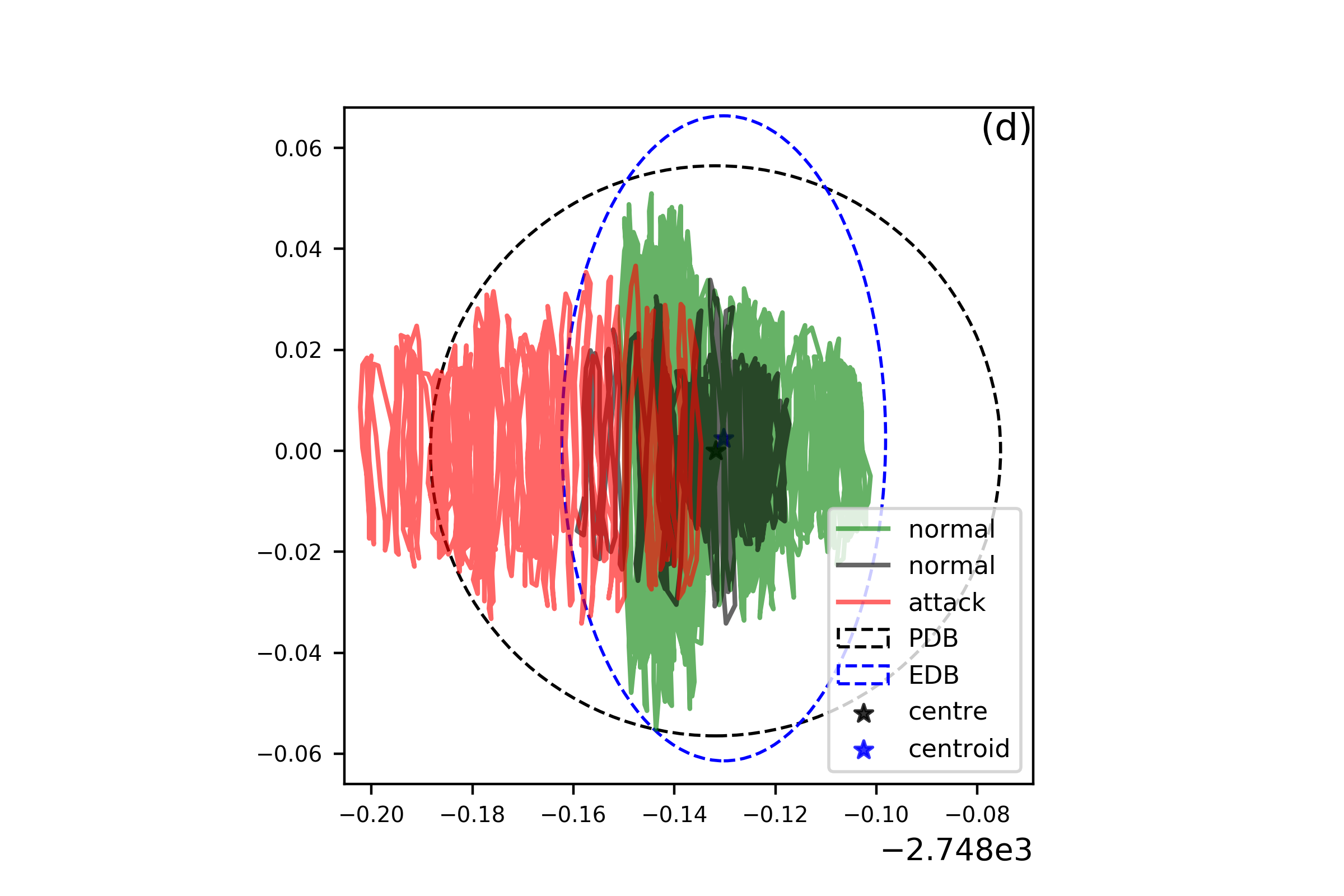}
	}\vspace{-.5cm}
	\caption{We demonstrate a stealthy attack scenario and its detection. Our proposed framework EPASAD is able to detect the attack more quickly than the baseline method PASAD. Subfigure (a) shows a sensor-generated measurements (by XMEAS(9) sensor, represents reactor's temperature). The green and black measurements are normal measurements used for training and validation, and the red measurements are captured under a stealthy attack (SA3). Subfigures (b) and (c) represent the departure score of corresponding measurements generated by PASAD and EPASAD frameworks. Subfigure (d) demonstrates the projections of each normal and attack measurement on the signal subspace (we consider a $2-$dimensional signal subspace for better visualization) and the decision boundaries of both, i.e., PASAD's decision boundary (PDB) and EPASAD's decision boundary (EDB).}
	\label{fig:SA3_8_2d}\vspace{-0.4cm}
\end{figure}

EPASAD uses the normal cluster to learn a uniformly tight and computationally efficient decision boundary. Many nonlinear functions such as convex/non-convex hull, skewed ellipsoid, higher-order nonlinear functions can envelop the signal subspace. Nonetheless, we use a specific ellipsoid function to parallel the standard axis of signal space to avoid any increase in online testing computation cost while ensuring a uniformly tight decision boundary for every dimension. We demonstrate EPASAD using Figure~\ref{fig:SA3_8_2d}. Figure~\ref{fig:SA3_8_2d}(a) represents the same attack scenario demonstrated in the Figure~\ref{fig:attackmodel1}. Figure~\ref{fig:SA3_8_2d}(d) shows an elliptical curve enveloping the 2-dimensional signal space within a minimum area. It brings the loose side of the decision boundary closer to the normal cluster, making each dimension uniformly tight. The elliptic decision boundary easily separates the abnormal red points that the spherical decision boundary misses. Hence EPASAD creates a challenging decision boundary for an attacker but is simpler to deploy. It does not give any redundant normal subspace where attacker can hide his abnormal activities.  


\subsection{Training of EPASAD framework} \label{subsec:training}

Consider a real-valued univariate time series $\mathcal{T}$=[$m_1$, $m_2$, $\cdots$, $m_{N}$,$\cdots,m_{N'}$, $ m_{N'+1}$,$\cdots$]. The subseries from $m_1$ to $m_{N}$ is used to determine $\mathbf{P}$=$\mathbf{U}^T$ while from $m_{N+1}$ to $m_{N'}$ as validation dataset. Before proceeding with the section, we list our assumptions.

\subsubsection*{Assumptions} \label{sec:assump} 

There are three basic assumptions to develop the EPASAD framework: \textit{(i)} the dataset used for training EPASAD can be noisy but cannot be anomalous. An anomalous pattern in training data can cause a data poisoning attack. \textit{(ii)} EPASAD is trained in an offline fashion, which needs all the training and validation datasets of length $N'$ to be available during training. \textit{(iii)} EPASAD prepares input features with the help of recent measurements that require an uninterrupted sequence of measurement.

\subsubsection*{Step 1: Generate normal cluster} \label{subsec:signalspace}

We collect the normal cluster by projecting the normal lagged vectors into the noise-free signal subspace. To determine $\mathbf{P}$=$\mathbf{U}^T$, EPASAD is trained over $\mathcal{T}[1$:$N]$ by utilizing the SSA and PASAD. The projection matrix projects an $L-$dimensional lagged vector from real space to an $R-$dimensional $(R\leq L)$ signal subspace. The projection matrix is trained over the series has a possibility of over-fitting the training data. Hence, we extend the normal training subseries with the validation datasets extending from $N$ to $N'$ ($N'>N$), i.e., ($\mathcal{T}[1$:$N']$). Thus, the trajectory matrix $\mathbf{M}$ for the extended validation subseries is of size $L\times K'$, where $K'$=$N'-L+1$ and each column vectors of $\mathbf{M}$ are projected to a signal matrix $\mathbf{X}$ of size $R\times K'$. The $i^{th}$ column vector is projected as $\mathbf{X}_i$=$\mathbf{U}^TM_i$. Hence, using Equation~\ref{eq:2signalnormalcluster} we project the entire $L-$dimensional matrix $\mathbf{M}$ to an $R-$dimensional signal matrix $\mathbf{X}$.
\begin{equation}\label{eq:2signalnormalcluster}
        \mathbf{X}=\mathbf{U}^T\mathbf{M}
\end{equation}
\subsubsection*{Step 2: Finding Centroid} \label{subsec:center}

We estimate the centroid $\hat{c}\in \mathbb{R}^R$ of the ellipsoid decision boundary using Equation~\ref{eq:3minmax}. Here, the elements of vector $\mathit{min}(\mathbf{X})$ are minimum elements of the corresponding dimension of $\mathbf{X}$ similarly, $\mathit{max}(\mathbf{X})$ are maximum elements. The mean of the cluster of a skewed sample distribution shift towards the dense side. Considering the mean as the centroid of the ellipsoid makes the decision boundary envelop the sparse side tightly and the opposite side loosely. Therefore, rather than choosing the projection of the mean of cluster to determine centroid as in PASAD, we determine the mid-point of the range of each dimension of $\mathbf{X}$. Further, we make centroid invariant signal subspace by using Equation~\ref{eq:4centroidinvariant} where $\mathcal{C}(x)$ is a centroid invariant element-wise squared vector. The centroid invariant signal subspace standardizes the ellipsoid decision boundary centered around zero-vector for every sensor.
\begin{equation}\label{eq:3minmax}
        \hat{c}=\dfrac{\mathit{min}(\mathbf{X})+\mathit{max}(\mathbf{X})}{2} 
\end{equation}
\vspace{-0.1cm}
\begin{equation}\label{eq:4centroidinvariant}
        \mathcal{C}(x)=(x-\hat{c})^2
\end{equation}
\vspace{-0.2cm}
\subsubsection*{Step 3: Learning Ellipsoid Decision Boundary} \label{subsec:ellips}

We determine the ellipsoid decision boundary that envelops the normal cluster in signal subspace $\mathbf{X}$. We consider a hypothesis function $f(x)$ for a variable vector $x\in \mathbb{R}^R$ to learn the decision boundary (cf. Equation~\ref{eq:5hypothesis}, here $w$ is a weight vector). When we express the hypotheses function $f(x)$ in the form of a standard ellipsoid function, the $\sqrt{w_i}$ describes the length of $i^{th}$ axis of the ellipsoid.
\begin{equation}\label{eq:5hypothesis}
    f(x)=w^T\mathcal{C}(x)
        =\dfrac{(x_1-\hat{c}_1)^2}{(w_1^{-0.5})^2}+\dfrac{(x_2-\hat{c}_2)^2}{(w_2^{-0.5})^2}+ \cdots+\dfrac{(x_r-\hat{c}_r)^2}{(w_r^{-0.5})^2}
\end{equation}

Our aim is to minimize the generalized $n-$dimensional volume to get minimum void space inside the decision boundary. Thus, we minimize the length of each ellipsoid axis such that all points of the normal cluster remain inside $f(x)$. Since the product of axis length is proportional to the ellipsoid volume, Equation~\ref{eq:6objective} is our objective function for learning the hypothesis function $f(x)$. Solving the objective function returns an optimal weight vector $\hat{w}$ that minimizes the product of the length of each axis ($\prod(w)^{-0.5}$). There are two hard constraints associated with the objective function \ref{eq:6objective}: \textit{(i)} $w^T\mathcal{C}(x)\leq 1$, forces each point to remain inside $f(x)$, and \textit{(ii)} $w>0$ assures an ellipsoid's real-valued axis length. We train the objective function over the column vectors of signal matrix $\mathbf{X}$ that gives an optimal weight vector $\hat{w}$ to get an optimal decision boundary. 
\begin{equation}\label{eq:6objective}
        \hat{w}=\arg\min_{w} \left (\prod(w)^{-0.5}\right ) | w^T\mathcal{C}(x)\leq 1,\;\forall x\in X\;\textbf{\&}\;w>0
\end{equation}

\subsubsection*{Step 4: Set Threshold}\label{subsec:threshold}

Since we train the objective function to minimize length of each axis of decision boundary using a hard constraint $w^T\mathcal{C}(x)\leq 1$, the value of $f(x)$ at a threshold $\theta_e$=$1$ is a decision boundary. The function $f(x)$ forms the tightest enveloping function $f(x)$, which does not consider any margin of error. However, a normal measurement can slightly deviate from the normal cluster causing false alarms. Thus, we add a margin of error, $\epsilon$,  also called slag-value in the threshold, $\theta_e$=$1+\epsilon$, to control the false alarms.

\subsection{Testing EPASAD Framework} \label{subsec:testing}
The EPASAD framework is deployed over SCADA to test each live streaming measurement in an online fashion. If $m_t$ is a measurement generated at timestamp $t$ and received by the SCADA, EPASAD prepares an $L$ length lagged vector $M_t$ using previous $L-1$ measurements; $M_t$=$[m_{t-L},m_{t-L+1},\cdots,m_t]^T$. The real-space lagged vector $M_t\in R^L$ are projected onto the $R$-dimensional signal subspace; $X_t$=$\mathbf{U}^{T}M_t$. For the most recent test measurement $m_t$, EPASAD computes a $\mathcal{D}_t$=$f(X_t)$. The $\mathcal{D}_t$ describes the confidence, regardless of whether the measurement is classified as an attack or normal. A smaller $\mathcal{D}_t$ indicates greater confidence of a measurement to be normal, while a higher $\mathcal{D}_t$ indicates greater confidence of an attack. A test measurement is classified as normal up to a tolerable value of the departure score threshold $\theta_e$. If $\mathcal{D}_t\geq\theta_e$, then EPASAD raises an attack alarm. This process completes the online testing step for a single measurement. The same procedure is repeated for the subsequent measurement generated at time $t+1$, and so on. The Algorithm~\ref{algo:test} depicts the pseudo-code of the EPASAD framework's online testing phase.

\SetCommentSty{mycommfont}

{\SetAlgoNoLine%
	\begin{algorithm}
		\SetKwInOut{Input}{input}\SetKwInOut{Output}{output}
		\Input{Lag parameter $L$, Dimensionality of signal subspace $R$ }
		\Output{An alarm when attack is detected} 
		\KwData{A test sequence $\mathcal{T}$}

		determine $\mathbf{U}$ \hspace{1cm}// Using SSA during training\\ 
		determine $c$ \hspace{1cm}// The centroid of ellipsoid\\
		determine $w$ \hspace{1cm}// By Equation~\ref{eq:6objective}\\ 
		$\theta_e\gets 1+\epsilon$ \hspace{1cm} //Set threshold\\
		
		\While(//Online testing the measurements streams){{(1)}}
		{
			$m\gets [m_{t-L},m_{t-L+1},\cdots,m_t]$\hspace{0.07cm} //Test sequence\\
			$x\gets U^Tm$ //Project $m$ to $R$ dimen. signal subspace\\
			$y\gets (x-c)^2$\\
			$\mathcal{D}_t\gets w^Ty$ \hspace{1cm}// Departure score for $m_t$\\
		    \If{$\mathcal{D}_t>\theta_e$}{
				   Raise an attack alarm\\
			    }
		}
		\caption{EPASAD's online testing}
		\label{algo:test}
	\end{algorithm}
}%

\subsection{Computation cost} \label{subsec:computation}

An IDS is deployed for the long term to secure the real-time streaming measurements from sensors. A sensor associated with ICS regularly sends measurements to the IDS; there may be a small-time difference between the streaming measurements. The IDS deployment must be efficient enough to generate the decision before proceeding to the subsequent measurement. Hence, online testing is crucial for low-cost hardware deployment. On the other hand, training is typically one time task accomplished in an offline fashion.

The main computation cost of EPASAD is the computing the departure score. The departure score evaluates a matrix to vector multiplication $x\gets \mathbf{U}^Tm$, it multiplies a $R\times L$ matrix to an $L-$dimensional vector requires $\mathcal{O}(RL)$ computing cost. Then, $y\gets (x-c)^2$ is an element-wise operation of two $R$-dimensional vectors with $\mathcal{O}(R)$ complexity. The final computation steps $D\gets w^Ty$ requires a dot product of two $R-$dimensional vectors, $\mathcal{O}(R)$. Hence, the overall computation cost of EPASAD is $\mathcal{O}(RL+R)$, which is equivalent to the computation cost of PASAD. Usually, only a few leading eigenvectors retain the majority of the signal information. Therefore, $R<<L$ is the average case of the computation cost. In the average case, the time complexity for online detection of EPASAD is linear in $L$, i.e., $O(L)$. 
The online deployment of EPASAD needs to store a projection matrix $\mathbf{U}^T$, centroid $c$, weight vector $w$, Which is require space to keep $RL$, $R$, and $R$ real numbers, respectively. Hence the space complexity of EPASAD is $\mathcal{O}(RL)$. Compare to PASAD, EPASAD needs to store an addition $R-$length weight vector $w$ which does not contribute much to space complexity. Hence both PASAD and EPASAD have the same space complexity of $\mathcal{O}(RL)$.

\section{Validation datasets} \label{sec:data}

We validate our proposed methodology using multiple attacks scenarios present in the two datasets listed below:
\subsection{The Tennessee-Eastman process dataset (TE-dataset)} \label{subsec:TEdata}

The TE-dataset is generated using an industrial chemical process simulation model proposed in 1993~\cite{downs1993plant}. The TE simulation framework mimics the process in a real-world chemical plant. The TE-process serves as a more realistic and safe environment for experimentation, transcending its original objective and becoming a popular choice among ICS security researchers~\cite{aoudi2018truth, zhu2017distributed, gao2016improved}. In ~\cite{aoudi2018truth}, the authors considered five attack scenarios to validate their method: three SAs and two DDAs. We consider two additional attack scenarios representing MSA and generate the TE-dataset by performing the following attacks.

\subsubsection{Micro-stealthy attack (MSA)}\label{subsubsec:MSA}

We consider two MSA attack scenarios to validate EPASAD. These include:

\begin{itemize}
\item \textbf{MSA1:} We simulate this attack by manipulating the process variable of a purge valve (XMV(6)). The XMV(6) restrict the reactor gas in the reactor tank from escaping into the atmosphere. Unnecessarily opening the valve more than a certain level causes low pressure in the reactor; Thereby causing the process to halt. Also, it causes unnecessary wastage of valuable gasses. In this scenario, we open the valve by 26\%, which is enough to degrade the system and waste the reactor gases but not that high to interrupt the process.

\item \textbf{MSA2:} We simulate this attack by manipulating the speed of an agitator (XMV(12)). The agitator ensures a well-mixed reactor, which impacts the heat transfer coefficients in the reactor. The maximum speed of the agitator should be 100\% to maximize the cooling capacity of the reactor coolant, and ideally, it is suggested to be 50\%~\cite{downs1993plant}. Hence, reducing the agitator speed below 50\% can increase the reactor's temperature, causing damages to the system. In this attack scenario, we consider the 38\% speed of the agitator which is slow enough to reduce the coolant capacity and increase the reactor's temperature. 
\end{itemize}

\subsubsection{Stealthy attack (SA)}\label{subsubsec:SA}

We consider three SA scenarios:
\begin{itemize}
\item \textbf{SA1:} We simulate this attack by manipulating the Stripper steam valve XMV(9). This valve controls the steam input to the stripping column. In this attack, we open the valve at 40\% compared to completely open.
\item \textbf{SA2:} We simulate this attack using the MSA1 attack scenario with a higher impact. In this attack scenario, we open the purge valve by 28\%, 2\% more than in MSA1.
\item \textbf{SA3:} We simulate this attack by tampering with the sensor XMEAS(10) to zero. The zero measurements of XMEAS(10) represent that purge valve XMV(6) is closed. For the counteraction, the controller would unnecessarily open the purge valve.
\end{itemize}

\subsubsection{Direct damage attack (DDA)}\label{subsubsec:DDA}

We consider two DDA scenarios:

\begin{itemize}
\item\textbf{DDA1:}  We simulate this attack by manipulating the process variable XMV(10) of a valve that controls cooling water flow to the reactor to prevent its temperature and pressure reach at a dangerous level. In this scenario, we open the valve to 35.9\%, which is lower than usual (41.106\%). Consequently, it increases the reactor's pressure and temperature and stops the process from reaching the maximum predefined limit.

\item \textbf{DDA2:} We tamper the reactor pressure sensor XMEAS(7) to zero. The controller takes action to perform more chemical reactions to maintain the reactor pressure. The abnormal increase in the pressure can damage the reactor, eventually stopping the process.
\end{itemize}

Each attack scenario of TE-dataset consists of measurements of 41 sensor as a time series. The dataset is collected for 48 hours, with the initial 40 hours under normal operation, and the remaining last 8 hours are during an active attack. The measurements are generated periodically such that it takes one hour to generate 100 measurements.
\subsection{C-town dataset} \label{subsec:batadal}

The C-town network dataset~\cite{taormina2018battle} is generated by simulating epanet CPA~\cite{taormina2017characterizing}. The network consists of 43 sensors and generates a measurement after every hour periodically. The dataset contains 14 distinct attacks launched in a different time window throughout nine months. The dataset contains three subdatasets, each of which consists of 43 process variables:
\begin{itemize}
\item \textbf{Subdataset 1}: It contains normal measurements during a period of one year.
\item \textbf{Subdataset 2}: It contains seven attacks along with normal operations during a period of six months.
\item \textbf{Subdataset 3}: It also contains seven attacks (but different) along with normal operations during a period of three months. 
\end{itemize}
Each subdataset, as mentioned above, is collected for the same sensor network. We combine subdataset 2 and 3 and call it subdataset 4 to evaluate EPASAD on the 14 attack scenarios captured during the nine-month-long period. 
The details of each attack scenario are provided in the paper~\cite{taormina2018battle}. 
\section{Experiments and results}\label{sec:experiment}

In this section, we validate our proposed method using above mentioned datasets and provide parameter values selected for the experiments.
\subsection{Experiment on TE-dataset}\label{subsec:exp1}
\begin{figure*}
	\minipage{\columnwidth} \vspace{-.5cm}
	\includegraphics[width=0.95\columnwidth]{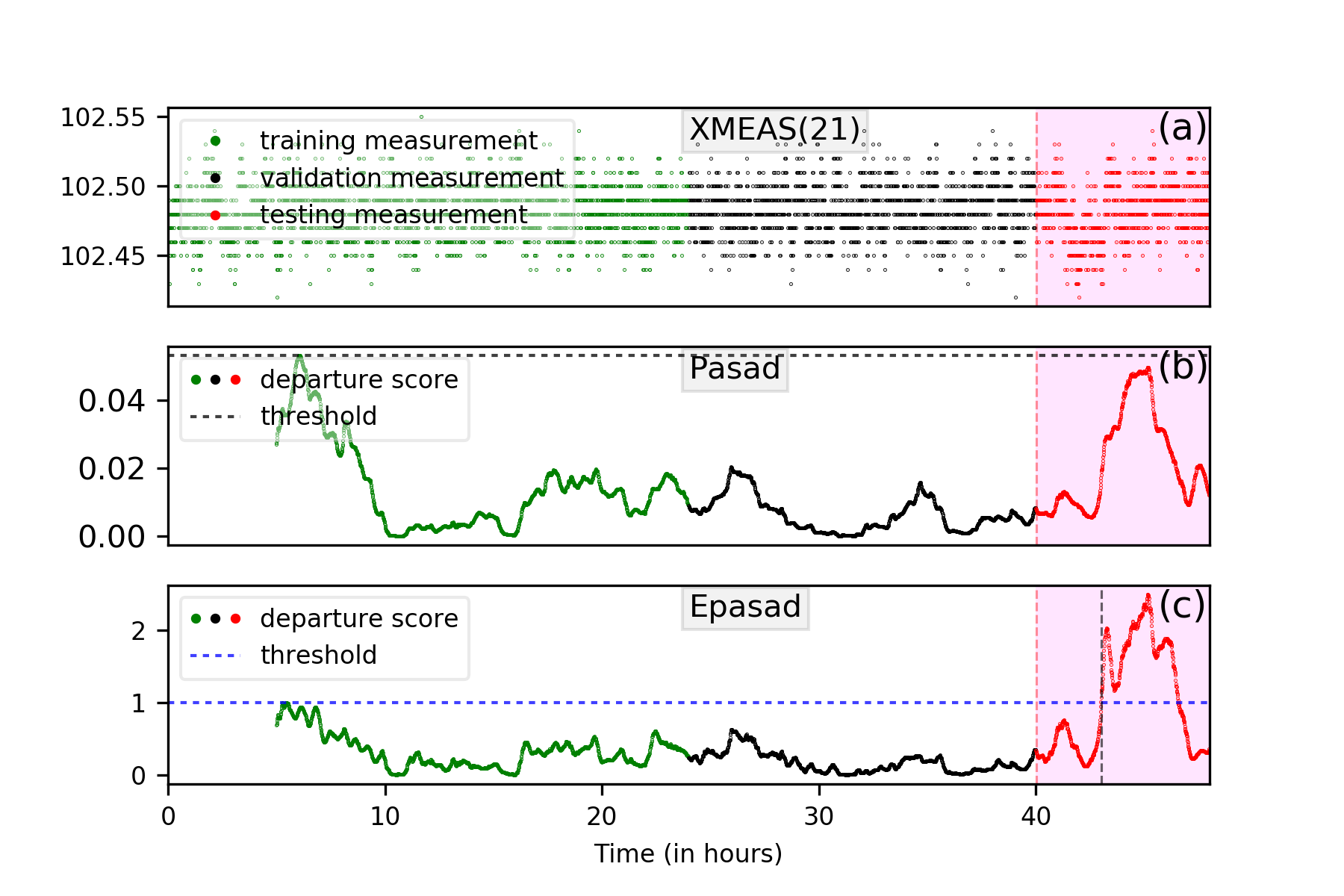}\vspace{-.5cm}
	\caption{We show the comparison of PASAD and EPASAD over sensor XMEAS(21) of TE-dataset. The attack measurements are collected during a micro-stealthy attack (MSA1) operation. EPASAD is able to detect the MSA, which PASAD fails to detect.}
	\label{fig:MSA}
	\endminipage\hfill
	\minipage{\columnwidth}\vspace{-.15cm}
	\includegraphics[width=0.95\columnwidth]{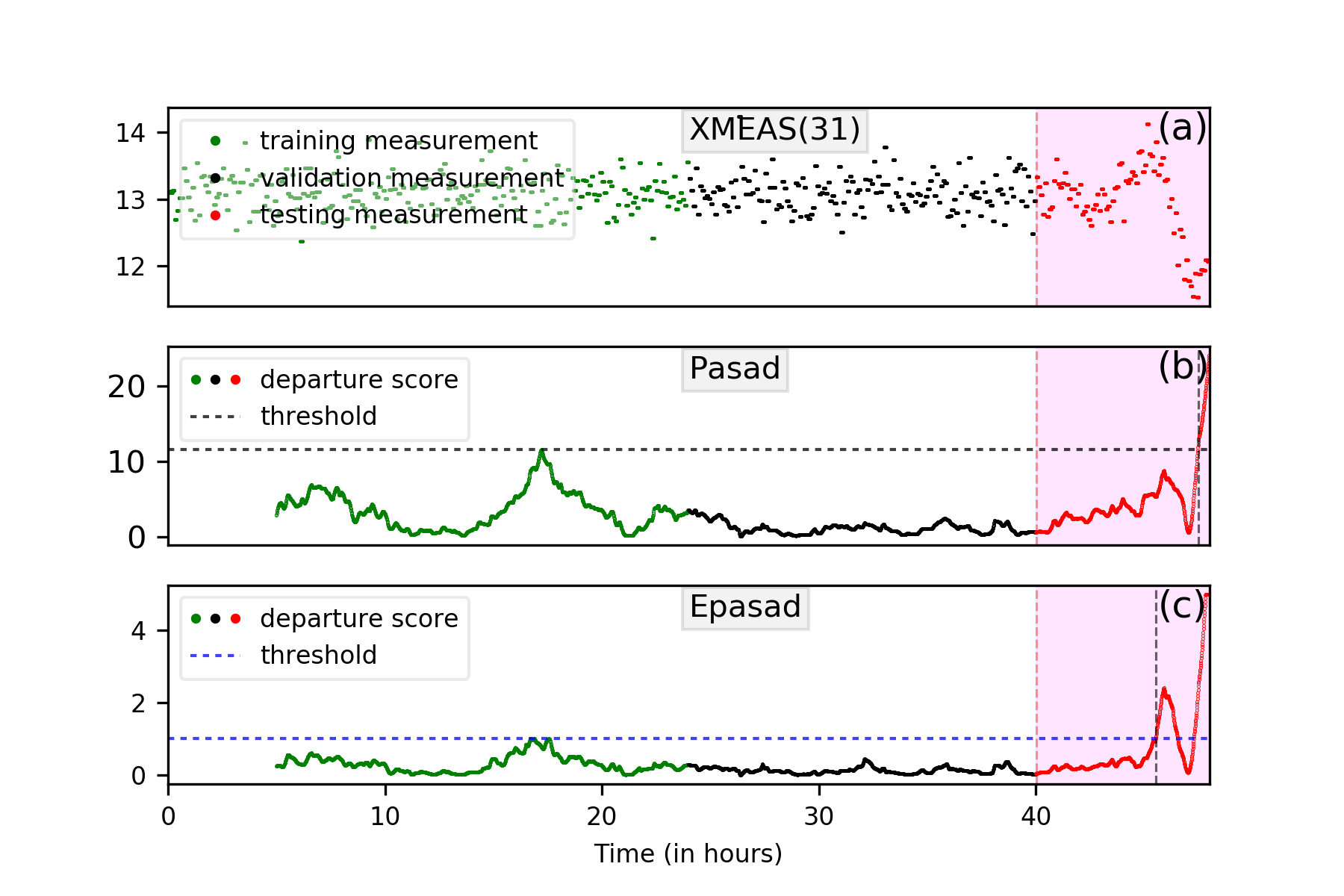}\vspace{-.5cm}
	\caption{We show the comparison of PASAD and EPASAD over sensor XMEAS(31) of TE-dataset. The attack measurements are collected during a Direct damage attack (DDA1) operation. EPASAD is able to detect the DDA more quickly.}
	\label{fig:DDA}
	\endminipage
\end{figure*}

In this experiment, we study how quickly we can detect SAs, MSAs, and DDAs. This experiment is carried out using comparable datasets and parameters for training, validation, and testing to make a fair comparison with the baseline method PASAD. Hence, we consider the normal subseries of the first 2400 (green) measurements to get the projection matrix $\mathbf{U}^T$ and then use the remaining 1600 (black) normal measurements as the validation dataset along with the training set to obtain the EPASAD decision boundary. We then apply the entire time series to EPASAD to do online testing. Figures~\ref{fig:SA3_8_2d},~\ref{fig:MSA}, and~\ref{fig:DDA} demonstrate the effectiveness of EPASAD towards detecting different attack scenarios in the TE process and comparing it with the baseline line method. 
Figures~\ref{fig:SA3_8_2d}(a),~\ref{fig:MSA}(a), and~\ref{fig:DDA}(a) represent the time series of sensor measurements. Figures~\ref{fig:SA3_8_2d}(b),~\ref{fig:MSA}(b), and~\ref{fig:DDA}(b) represent the corresponding departure score by applying the baseline method PASAD. Figure~\ref{fig:SA3_8_2d}(c),~\ref{fig:MSA}(c), and~\ref{fig:DDA}(c) represents the departure score by applying our proposing method EPASAD. Similar to~\cite{aoudi2018truth}, we also set the threshold at maximum departure score of the normal measurements hence there was no false-alarm in the TE-dataset scenarios. Therefore, all the evaluation of this dataset is represented in term of recall only.

\begin{figure}
\centering
\includegraphics[width=0.95\columnwidth]{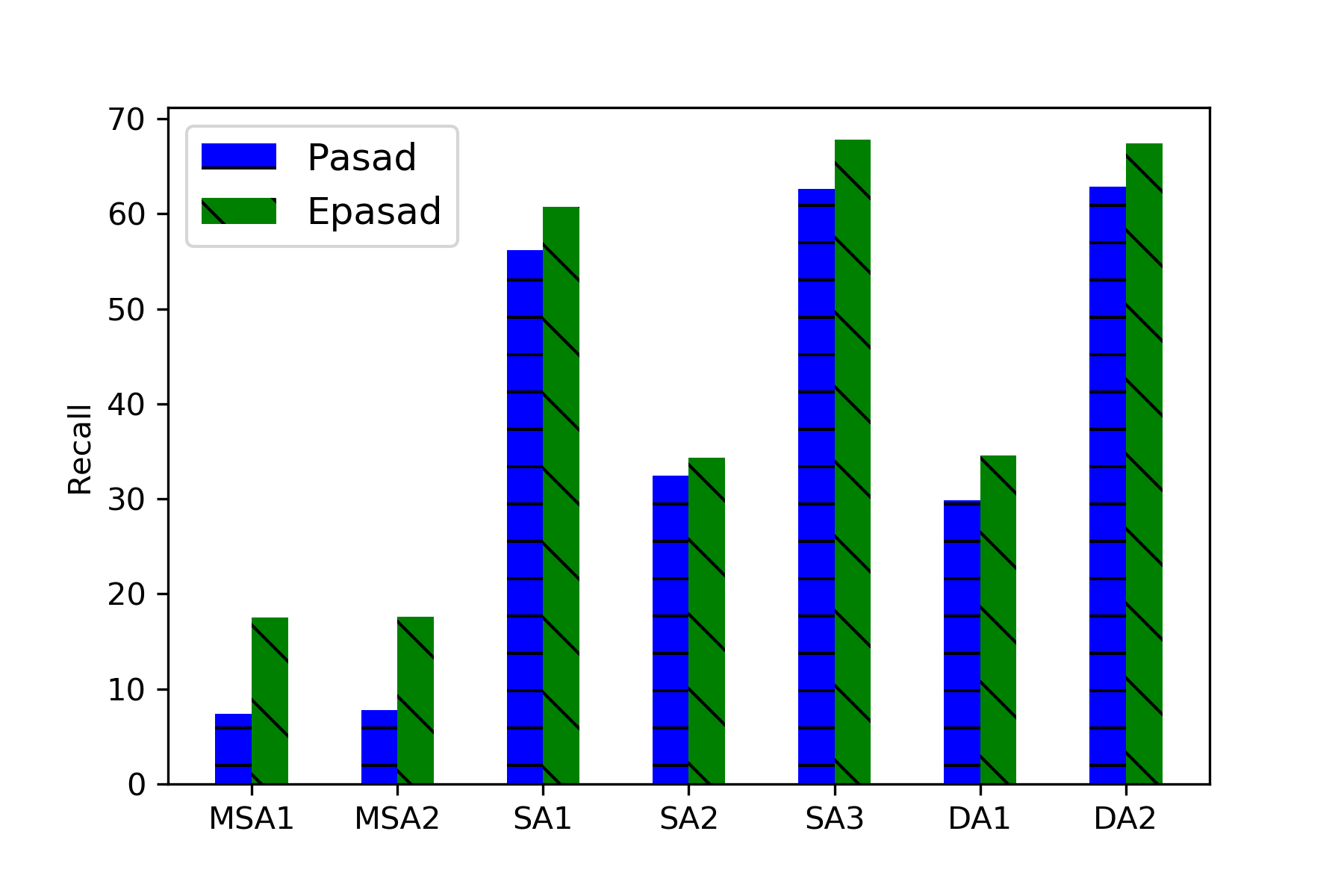}\vspace{-.5cm}
\caption{We compare the accuracy of PASAD and EPASAD over the seven different attack scenarios of the TE-dataset in terms of recall. EPASAD achieves a significant improvement in each attack scenario.} \label{fig:all7recallTE}
\end{figure}

Figure~\ref{fig:SA3_8_2d} shows sensor operating under SA scenario. The part of the subseries that has been captured under SA appears to be normal. Such anomalous series when projected on the signal subspace are significantly far from the normal cluster. PASAD's departure score takes a long time to be more than $\theta_p$, causing a delay in detecting the attack. Moreover, the departure score raising alarm returns to normal after a short period, which an administrator may think of as a false alarm. On the other hand, EPASAD detects the attack shortly after it begins and raises the alarm for an extended time. Hence, \textbf{EPASAD is more effective at detecting SAs quickly}. Further, we evaluate EPASAD on each process variable of SA scenarios SA1, SA2, and SA3. Our results (cf. Figure~\ref{fig:all7recallTE}) show a significant improvement in all the attack scenarios. EPASAD improves the average recall of all three SAs from 50.3\% to 54.2\% compared to the baseline benchmark.

We demonstrate our method on a process variable which is captured under MSA (cf. Figure~\ref{fig:MSA}). The results show that the departure score of PASAD is always less than the $\theta_p$ during the attack. Hence, it could not detect MSA. On the other hand, EPASAD computes a significant departure score which is more than the $\theta_e$ for a lengthy period. Hence, \textbf{EPASAD is able to detect even the MSA}. We tested EPASAD on every process variable in the MSA1 and MSA2 datasets. The results (cf. Figure~\ref{fig:all7recallTE}) show significant improvement with the average recall increasing from 7.5\% to 17.3\%.

We evaluate our method on a process variable of the DDA1 attack scenario (cf. Figure~\ref{fig:DDA}). In this scenario, the measurements during the attack operation are initially close to normal and then suddenly become abnormal, even beyond the normal range (the lower and upper limit of measurements generated by a sensor). The baseline method PASAD could not recognize the initial symptoms. It detects the attack when the attack induced-measurements reach beyond the normal range. On the other hand, EPASAD detects such attacks at early stages, shows a significant gain over the baseline method. Hence, \textbf{EPASAD can quickly detect the DDAs}. Figure~\ref{fig:all7recallTE} shows the average performance of EPASAD on each process variable of the DDA1 and DDA2 attack scenario. Here, EPASAD improves recall score from 46.2\% to 51.0\%.

\begin{figure}
\includegraphics[width=0.95\columnwidth]{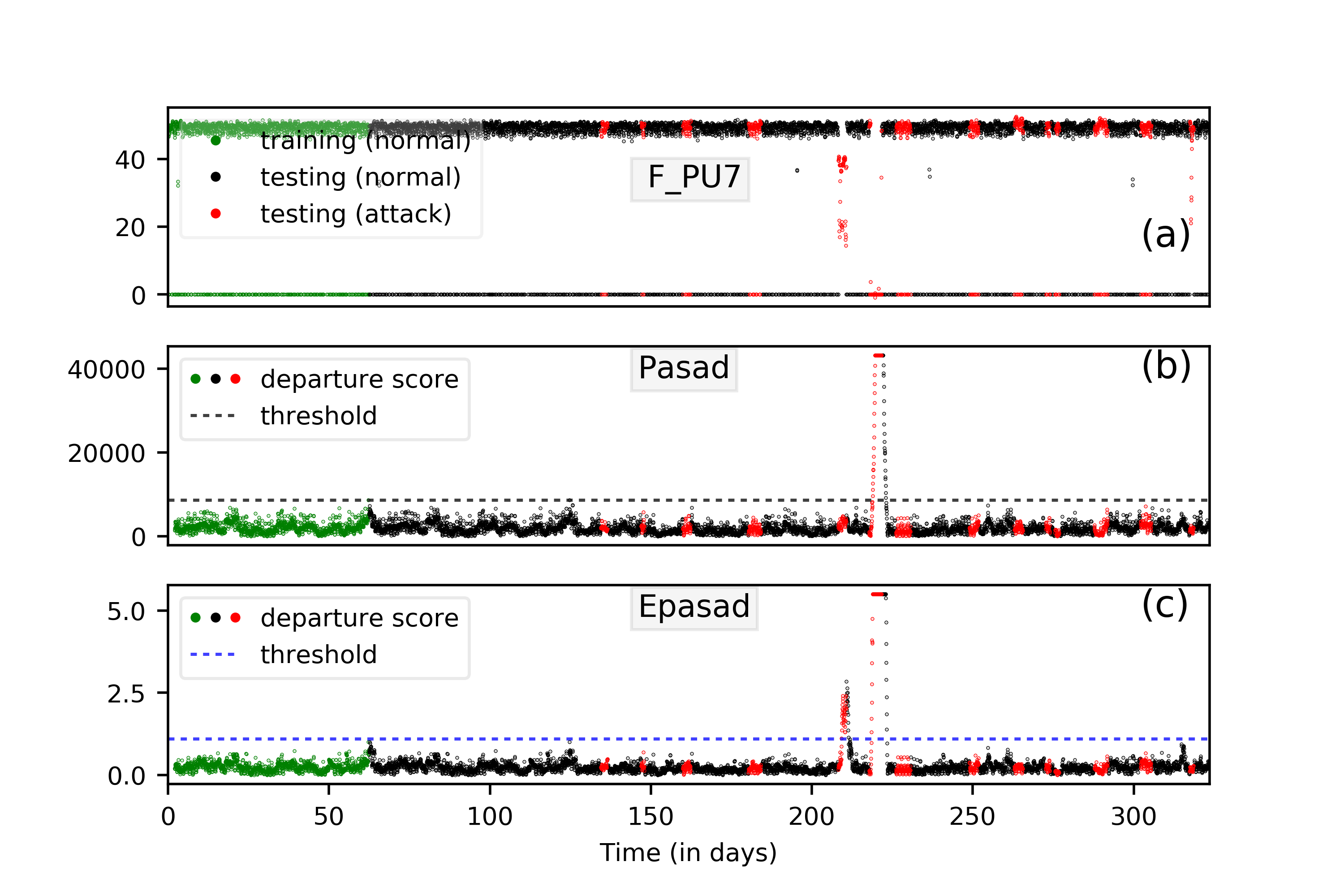}\vspace{-.5cm}
\caption{Comparison of PASAD and EPASAD over the sensor $FPU7$ of C-town dataset, collected during 14 different attacks where PASAD fails to detect the abnormality induced during the $5^{th}$ attack, and EPASAD is able to detect it. The green measurements are normal measurements used for training. The black and red measurements are normal, and attack measurements are used for testing. Note that the order of each subfigure has the same definition as Figure \ref{fig:MSA} and \ref{fig:DDA}.}
\label{fig:B_20}
\end{figure}
\subsection{Experiment on C-town dataset} \label{subsec:exp2}

In a realistic scenarios, attacks are launched for a limited duration, and then the system resumes normal operation. The 14 attacks in this experiment are launched for a limited time before the system resumes normal operation. This is recurrent and done over a period of 9 months. Figure~\ref{fig:B_20} demonstrates EPASAD on a process variable of the C-town dataset. We train EPASAD over a subseries of length 1500 (green measurement) captured under normal operation to get the projection matrix. Then, include 1500 normal measurements (black measurements next to the green ones) as validation dataset to determine the decision boundary. Once the training phase is complete, we test the entire subseries using the online testing algorithm~\ref{algo:test}. Figure~\ref{fig:B_20}(c) indicates EPASAD's strengths in detecting the structural changes caused by the $5^{th}$ and $6^{th}$ attacks reflected in the $FPU7$ sensor and then return to the normal state.

In Tables~\ref{tab2} and~\ref{tab3}, we evaluate the experiment at the entire infrastructure level by aggregating the nature of alarms in every process variable. If the IDS triggers an alarm in any processes during an attack, we consider the attack to be detected. We consider a false alarm if it is triggered in any sensor during the normal operation. Table~\ref{tab2} evaluates each attack using two attributes, time (in hours) and count. The time field represents how long an attack has been active without causing an alarm to be raised. In other words, it is the time taken by IDS to raise the first alarm. The count field represents the number of process variables involved in the alarm's triggering. It is very unusual to raise false alarms in multiple sensors at a time, a higher number of counts sensors producing attack alarm increases the confidence of positive alarm. Table~\ref{tab3} evaluates the overall accuracy in terms of true alarm rate (recall), precision, F1-score, and false alarm rate.

\begin{table*}
\caption{Performance and comparison of PASAD and EPASAD framework for all 14 attacks present in the C-town dataset. The comparisons are based on the time (in hours) taken to detect an attack and the number of sensors that trigger the alarms. Here `$\times$' represents an attack not detected.}\label{tab2}
\centering
\begin{tabular}{|l|l|l|l|l|l|l|l|l|l|l|l|l|l|l|}
\hline
Attack &  1 & 2 & 3 & 4 & 5 & 6 & 7 & 8 & 9 & 10 & 11 & 12 & 13 & 14\\
\hline
Time (PASAD) &  0 & $\times$ &  4 & 17 & 10 & 14 & 26 & $\times$ & 22 &  0 & 10 & 52 &  4 & 17\\
Time (EPASAD) &  0& 10&  3& 16& 12& 16& 32& 18& 0&  0& 10& 18&  4& 10\\
\hline
Count (PASAD) &  6&  0& 9&  6&  3&  5&  5&  0&  4&  6& 11&  3&  3&  2\\
Count (EPASAD) &  6&  2& 10&  6&  6&  5&  5&  3&  8&  7&  9&  6&  4&  2\\
\hline
\end{tabular}
\end{table*}

\begin{table}
\caption{The average performance and comparison (in percentage) of PASAD and EPASAD frameworks on the C-town dataset}\label{tab3}
\centering
\begin{tabular}{|l|l|l|l|l|}
\hline
   & Precision & Recall & F1-score & False Alarm\\
\hline
PASAD &  64.36 & 54.84 &  59.22 & 4.36\\
\hline
EPASAD &  71.36& 64.29&  67.64 & 3.70\\
\hline
\end{tabular}
\end{table}

This experiment tests the long duration when measurements are captured under mostly normal operation and sometimes under various attacks. Hence, there is a possibility that an IDS in this experiment generates a large number of false alarms. The results in Table~\ref{tab3} show a significant improvement in the recall score (true alarm rate) and a low false alarm rate. In addition to the overall performance, we analyze the detection of all 14 attacks in Table~\ref{tab2}. We analyse the time taken to detect an attack and the count of the number of sensors engaged in triggering an alarm. EPASAD has a significant gain in detecting the two attacks ($9^{th}$, and $12^{th}$) over PASAD, and EPASAD even detects the two missing ($2^{nd}$ and $8^{th}$) attacks. EPASAD generates a valid alarm in more number of sensors that increase the alarm's confidence. Hence, \textbf{EPASAD can quickly and confidently raise the alarm for detecting an attack}. 
EPASAD slightly under-performs in three scenarios (cf. $5^{th}$, $6^{th}$, and $7^{th}$ attack scenario in table \ref{tab2}) of the C-town dataset. In PASAD, if the projection is on the tight dimension, it performs slightly better. EPASAD slightly loosens each dimension by adding a small value ``slag'' to the threshold. Thus, if the projection is in the tight dimension, PASAD might be better. But in general, keeping the attack such that all dimensions are tight is hard for the attacker to find the loosest side as they would now need to identify radii in each dimension.

\subsection{Parameter selection} \label{subsec:exp3}

In this section, we discuss the parameters and their choices that help us in implementing the above experiments. We use the same datasets and parameters to experiment with PASAD and EPASAD to make a fair comparison. There are two main parameters that are required in the training phase: lag $L$ and dimensionality of signal subspace $R$.  The lack of generalization of parameters in the baseline paper encourages us to choose the best performing parameter for PASAD. We run PASAD over various lag values, from 100 to 1000 in the increments of 100 for the TE dataset and 20 to 200 in increments of 10 for the C-town dataset to find the best lag value. We find the best performing lag parameter, $L$=$500$ for TE-dataset and $L$=$50$ for the C-town dataset. A smaller value of the lag parameter for the C-town dataset yield the best results because the time between two consecutive measurement is one hour, while the TE-dataset generates 100 measurements in one hour. Hence, a subsequence of length 50 itself covers the subsequence of more than two days. The dimensionality of signal space $R$=$3$ is found to be best performing. Once the training is finished, we set a threshold $\theta_e$ to classify the departure of measurement between attack and normal. The experiment~\ref{subsec:exp1} of the TE dataset uses entire normal subseries for training and validation, which ensure no false alarm with a minimum threshold with slag-value $\epsilon$=$0$. In the experiment~\ref{subsec:exp2}, when we set $\theta_p$ to the maximum of validation subseries without adding any slag-value, we find that PASAD fails to detect two attacks ($2^{nd}$ and $8^{th}$). Adding a slag-value could fail to detect more attacks and decreases the alarm. On the other hand, EPASAD is tighter in each dimension has a higher chance of raising a false alarm. Hence, we add a slag-value $\epsilon$=$0.1$ in $\theta_e$ to ensure a lesser false alarm rate.

\section{Related works} \label{sec:related}
In this section, we discuss earlier IDSs in the industrial control system. In~\cite{aoudi2018truth}, the authors published a method to detect attacks in ICS at a process variable label named PASAD. PASAD is a univariate departure-based process-level detection method that can detect even a SA on control systems by identifying an abnormal sequence. There are two other popular process level detection methods: Linear Dynamic State-space (LDS) by~\cite{shoukry2015pycra} and the Auto-Regressive (AR) methods~\cite{hadvziosmanovic2014through} (which we describe later in the section). A comprehensive survey of these methods is presented in~\cite{urbina2016survey}.

Along with the univariate process-level detectors, there are other popular multi-process-level detectors methods. In~\cite{guan2003means}, the authors used the K-Means clustering method along with the algorithms discussed in~\cite{hansen2001j} and named it Y-Mean clustering method for network intrusion detection. This method is tested on the KDD99 dataset. In~\cite{hu2008adaboost}, the authors applied the AdaBoost algorithm on the KDD99 dataset and achieved better accuracy with fewer false alarms.
Further, different studies also use reconstruction-based deep learning methods~\cite{feng2017multi,goh2017anomaly, taormina2018deep}. In~\cite{feng2017multi}, the authors combined the Long Short-Term Memory (LSTM) network with a bloom filter to detect the malicious traffic in the gas pipeline SCADA dataset. In~\cite{goh2017anomaly}, the authors predicted the next measurement using the LSTM and checked both positive and negative deviation from actual measurement, validating the method on water treatment testbed datasets. Similarly, in~\cite{taormina2018deep}, the authors used the AutoEncoder model to reconstruct a measurement, and if it is found a higher deviation from the actual, then trigger an alarm. The method is further improved by using cumulative sum (CUSUM). 

A process-level IDS is categorised in two categories, the univariate (independent IDS for each sensor variables)~\cite{aoudi2018truth,shoukry2015pycra,hadvziosmanovic2014through,aoudi2021framework} and multivariate (an IDS model takes input from the multiple sensor variables)~\cite{guan2003means, hansen2001j, hu2008adaboost, nader2014l_p, feng2017multi, goh2017anomaly, taormina2018deep, aoudi2020scalable}. In~\cite{garcia2017hey}, the authors developed a PLC rootkit that can corrupt the communication route between sensors and SCADA. An attacker can compromise a few communication channel and manipulate them accordingly to misclassify the structural changes in any other sensors as well. In~\cite{erba2020constrained}, the authors used this concept to construct an evasion attack against multivariate detectors~\cite{feng2017multi,goh2017anomaly, taormina2018deep}. On other hand, an univariate detectors are independent model for each sensor. Manipulating a few sensor  measurements cannot evade any other univariate IDS model.

There are four univariate process-level-based detectors methods: LDS methods, AR methods, PASAD and PADS. In~\cite{urbina2016survey}, the authors survey and explain a model that uses the LDS method with a time delay to detect the pH water level using SWaT testbed~\cite{mathur2016swat}. In~\cite{cardenas2011attacks}, the authors created several TE process attacks and used LDS together with non-parametric CUSUM statistics. In~\cite{shoukry2015pycra}, the authors used the model together with $\chi^2$ anomaly detection technique to extend it for various kinds of sensor variables named it PyCRA. These LDS-based methods are challenging to build. They need a detailed description of process variable that may not always be available~\cite{feng2017multi,kiss2015clustering}. In~\cite{hadvziosmanovic2014through}, the authors leveraged auto-regressive model with Shewhart control limits on time series extracted from the Modbus PLC traffic, evaluated their approach on two water treatment testbed datasets. The result of this method is compared with the PASAD in~\cite{aoudi2018truth}. The authors found that the AR model fails to detect the SAs and delay detecting the DDA; hence PASAD is found more substantial to detect those attacks. In~\cite{aoudi2021framework}, the authors present another univariate framework called PADS, which uses departure score of PASAD to classify an alarm in two categories, weak alarm and actionable alarm using two thresholds setting. This framework determines a higher threshold that classifies the alert as an actionable alert. It reduces the frequency of false alarms also recall. Similarly, for weak alert, it increases the false alarm rate as well as recall. Hence, it is difficult to compare the results with this framework. Since EPASAD is improving the departure score of PASAD can improve PADS as well.

In summary, we find four process-level detector methods~\cite{aoudi2018truth,shoukry2015pycra,hadvziosmanovic2014through,aoudi2021framework} where PASAD is the most accurate univariate process-level data-driven method to detect attacks in critical infrastructures, therefor we consider PASAD for baseline comparison. Our proposed method EPASAD improves the performance without hurting its any strengths. The detailed comparison of EPASAD with PASAD by using two popular benchmark shows that the proposed method EPASAD is more accurate than PASAD, and it detects attacks that PASAD fails to detect.

\section{Discussion and Conclusion}\label{sec:discandconclusion}

The CIs are vulnerable to cyber-attacks, primarily due to the importance of CIs to the nation and society. In a world full of threats, attackers successfully breach the many tiers of CI security. This research presents a last-layer security solution called EPASAD framework to detect an attack after an attacker has successfully evaded all network security and begun harming the CIs. EPASAD is a univariate, light-weighted, process-level, non-parametric, data-driven, and model-free attack detection framework, that is motivated to detect even tiny structural changes hidden within the noise margin of a process variable. To validate the EPASAD framework, we introduce a MSA scenario, which is extremely difficult to detect by any available methods, but EPASAD efficiently detects it. EPASAD detects quickly every other attack scenario considered for validation and significantly improves the performance of PASAD without any additional computational overhead. We summarise the following six essential strengths of EPASAD based on our experiments on various attack scenarios and available literature: 

\begin{itemize}
\item \textbf{EPASAD quickly detects an attack:} EPASAD aims to detect even tiny structural changes in the normal behaviour of the sensor and detect even MSA attack at the very initial stages (cf. Figure \ref{fig:MSA}). Based on the experiments performed, EPASAD improves the performance of detecting the attacks in all attack scenarios, including seven of TE-dataset and fourteen of C-town dataset  (cf. Section~\ref{sec:experiment}). 
In a most unlikely scenario,  when the signal space is equally distributed across each dimensions, EPASAD can still learn a uniformly tight n-spherical decision boundary. Thus, EPASAD's performance will always be better than PASAD.

\item \textbf{EPASAD also works under noisy environment:} In~\cite{mo2015performance}, the authors highlighted the critical problem of making the unrealistic assumption that the system model is noiseless. A noisy environment can cause severe problems for a non-robust IDS. An attacker can hide their malicious manipulations within the noise, and the noisy environment causes lots of false alarms. Our proposed method, EPASAD, is based on a well-known robust time series tool called SSA. The SSA is suitable to capture the skeleton of deterministic pattern from a noisy time series that makes EPASAD robust enough to work even in a noisy environment (cf. Chapter 6 of~\cite{elsner2013singular}). 

\item \textbf{EPASAD is realistic to build and deploy:} EPASAD is a non-parametric and purely data-driven framework that does not need prior knowledge of the system or the family of the probability distribution of the time series data. Hence we have not used any prior knowledge of sensors measurement distribution to model EPASAD in our experiment (cf. Section~\ref{sec:experiment}).

\item \textbf{EPASAD is computationally efficient:} EPASAD is developed to deploy over real-time CI, which requires processing the streaming measurement. EPASAD is a lightweight framework that produces a decision for measurement in linear time complexity of $O(L)$ in order of lagged vector. EPASAD is tested on a `Intel(R) Core(TM) i7-4770 CPU @~3.40GHz' machine with `64-bit Ubuntu 16.04 LTS' operating system and `16 GB' RAM. EPASAD takes 3.6 and 3.0 $\mu$sec to generate one result for TE-dataset and C-town datasets, respectively. 

\item \textbf{EPASAD is secure against evasion attack:} In~\cite{garcia2017hey}, the authors developed a PLC rootkit that can corrupt the communication route between sensors and SCADA. An attacker can compromise a few communication channels and manipulate them accordingly to hide the structural changes in the normal behaviour of any other sensors. In~\cite{erba2020constrained}, the authors used this concept to construct an evasion attack against multivariate detectors~\cite{feng2017multi,goh2017anomaly, taormina2018deep}. In the case of univariate IDS, each sensor is independently modeled. Manipulating a few sensor variables cannot affect any other univariate IDS model. Hence univariate IDS are safer against evasion attacks.

\item \textbf{EPASAD generates a low false alarm rate:} unlike any other nonuniform decision boundary-based model in which low margin sides are volatile to raise a false alarm. EPASAD is motivated to learn a uniform decision boundary, and adding a small slag-value provides a margin of error without compromising accuracy. As a result, EPASAD generated only $3.70\%$ false alarm (cf. Table~\ref{tab3}) while testing it for nine months.
\end{itemize}

Identifying the structural changes in time series data is a classical problem that is useful for detecting irregularities and attacks in a wide range of applications such as an automated vehicle, robotics, UAVs, IoT, etc. Improving the performance of detecting the structural changes in a time series data can also enhance the other applications that will be developed in the future. In addition to using EPASAD in other domains, we would like to extend it as a multivariate model, which can be computationally more suitable for large sensor-connected networks.

\section*{Acknowledgement}

This work is partially funded by the C3i Center's funding from the Science and Engineering Research Board of the Government of India. 

\bibliographystyle{ieeetr}
\bibliography{bibfile.bib}

\end{document}